%
%
%

\input harvmac
\def\heading{\newsec}
\def\subheading{\subsec}
\newdimen\cboxindent  
\newdimen\cboxflex    
\cboxindent=3.5cm
\cboxflex=5cm
\def\cbox#1{\vbox
   {\advance\leftskip by\cboxindent plus\cboxflex minus\cboxflex
    \advance\rightskip by\cboxindent plus\cboxflex minus\cboxflex
    \parindent=0pt
    \parfillskip=0pt
\hyphenpenalty=10000
    #1}}
\def\tiny{\scriptscriptstyle}
\def\tablerule{\noalign{\hrule}}
\def\smallstrut{\vrule height5pt width0pt depth 0pt}
\def\deepstrut{\vrule width0pt depth 10pt}
\def\strutrule{\vrule height12pt width0pt depth 5pt}

\cbox{\bf SPECTRAL DENSITY STUDY OF THE SU(3) DECONFINING PHASE TRANSITION$^{\dag}$}
\vfootnote\dag{Submitted to Nuclear Physics B}
\bigskip\bigskip

\centerline{Nelson A. Alves$^{\tiny 1}$, Bernd A. Berg$^{\tiny{1,3}}$
\footnote{$^{**}$}{On leave of absence from Department of Physics,
The Florida State University, Tallahassee.}
~and Sergiu Sanielevici$^{\tiny{2,3}}$}
\bigskip
\bigskip
\centerline{$^{\tiny 1}$Fakult\"at f\"ur Physik}
\centerline{Universit\"at Bielefeld, D-4800 Bielefeld 1, Germany}
\bigskip

\centerline{$^{\tiny 2}$Department of Physics}
\centerline{The Florida State University, Tallahassee, FL 32306, USA}
\bigskip

\centerline{$^{\tiny 3}$Supercomputer Computations Research Institute}
\centerline{The Florida State University, Tallahassee, FL 32306, USA}

\vskip2in
{\smallskip\narrower
\centerline{\bf Abstract} 
\bigskip

We present spectral density reweighting techniques adapted to
the analysis of a time series of data with a continuous range of 
allowed values. In a first application we analyze action 
and Polyakov line data from a 
Monte Carlo simulation on $L_t L^3\ (L_t=2,4)$ lattices for the SU(3) 
deconfining phase transition. We calculate partition function zeros,
as well as maxima of the specific heat and of the order parameter 
susceptibility. Details
and warnings are given concerning i) autocorrelations in computer
time and ii) a reliable extraction of partition function zeros.
The finite size scaling analysis of these data leads to precise
results for the critical couplings $\beta_c$, for the critical 
exponent $\nu$ and for the latent heat $\triangle s$. In
both cases ($L_t=2$ and 4), the first order nature of the 
transition is substantiated.
\smallskip}
\vfill\eject

\heading{INTRODUCTION} 

Monte Carlo simulations of finite statistical systems at a coupling
$\beta=\beta_0$ generate a time series of correlated data such
that for appropriate observables ${f}$ the arithmetic average
of measurements

$$ 
{\overline{f}} (\beta_0)\ =\ {1\over N} \sum_n^N {f}_n
\eqno(1.1) 
$$
becomes an estimator of the expectation value $<{f}> (\beta_0)$:

$$ 
{\hat {f}} (\beta_0 )\ =\ <{f}> (\beta_0)\ 
=\ \lim_{N\to\infty} {1\over N} \sum_n^N {f}_n . \eqno(1.2) 
$$
Using the spectral density representation of the partition function

$$ Z(\beta)\ =\ \int_{S_{\min}}^{S_{\max}} dS\ n(S)\ \exp (\beta S) ,
\eqno(1.3) 
$$
reweighting techniques allow to calculate the estimator
${\overline{f}} (\beta)$ for all $\beta$ in a sufficiently small
neighborhood of $\beta_0$. Although reweighting techniques have a
long history [1-9], the extent of practical improvements became only
fully realized after the recent work by Ferrenberg and Swendsen [7,8].
In this paper we further elaborate on these techniques and cast them
in a form suitable for practical applications in lattice gauge 
theories, where one has to deal with a time series of continuous
variables. In particular, procedures for
combining (``patching'') data from simulations at various 
$\beta_0$ values are discussed in some 
detail. Altogether, we have in mind to establish a model analysis
which may provide useful guidance for the analysis of data from
future large scale simulations of the QCD deconfining phase transition,
like the U.S. teraflop project~[11]. Here, we give a reweighting 
analysis for the SU(3) deconfining phase transition. In a subsequent 
paper we shall present our SU(2) results. (From the point of view of 
reweighting techniques, the SU(2) theory turns out to be the more 
difficult case, because the action density distributions are barely
distinguishable from Gaussian distributions.)

Pioneering numerical work on the SU(3) deconfining transition was
done some years ago [12]. 
Renewed interest was stimulated by the APE collaboration [13], who raised 
doubts about the first order nature of the transition. This prompted
a number of finite size scaling (FSS) studies [14-17] with the result 
that the first order nature of the transition was rather unambiguously
established. Here, we give the details promised in [17]. Beyond~[17]
we present an investigation of autocorrelation times and a theoretical
analysis concerning the numerical calculation of partition function
zeros. The latter point enforces some corrections of previously stated
results. Aside from quantities which are functions of the action 
(energy)~[17], we also analyze now the Polyakov line susceptibility.

The paper is organized as follows: Section~2 introduces the spectral
density method as used in this paper, section~3 investigates 
autocorrelations in computer time (for a review see [18]) and makes a 
connection to error bar calculations by binning [19,20], section~4 
gives our reweighting calculations for the specific heat and a FSS estimate 
for the latent heat, section~5 is devoted to our  calculation of 
partition function zeros and their FSS analysis. Beyond earlier 
approaches~[5,21] a consistency argument is developed. In close analogy
with work in~[15] we give in section~6 a reweighting and FSS analysis for 
the Polyakov loop susceptibility. Summary and conclusions are contained 
in a final section~7.
 
\medskip
\heading{SPECTRAL DENSITY MC CALCULATIONS } 

We consider the SU(3) Wilson action

$$ 
S = \sum_p S_p ~~~{\rm with}~~~ 
S_p = {1\over 3} Tr (U_p) . \eqno(2.1) 
$$
where the sum goes over all plaquettes of the four-dimensional lattice
with volume $V=L_t L^3$ . The MC simulation provides us with a time
series of measurements for the action: $S_n,\ (n = 0,...,N)$ where
$N$ is the total number of measurements. We generate our data by means
of the computer program published in ref.~[22], taking 
measurements (meas.) after every sweep. A certain number
of initial sweeps are omitted for thermalization (therm.). To have a 
convenient normalization, we monitor the corresponding action per 
plaquette (not to be confused with $S_p$)

$$ 
s_n = {S_n \over V_p}, \eqno(2.2) 
$$
where $V_p=6V$ is the total number of plaquettes.
Let us first consider a single MC simulation at a fixed value 
$\beta = \beta_0$. What is the $\beta$ range of validity for a spectral 
density as obtained from the MC simulation at $\beta_0$? Let us assume 
it is valid for $\beta \in [\beta_{\rm min}, \beta_{\rm max}]$. 
In this range we are able 
to calculate estimators for quantities of physical interest by

$$ 
{\overline f} (\beta ) = {F \over Z} ~~~{\rm with}~~~ 
F = \sum_{n=1}^N f_n \exp \left( \triangle \beta S_n \right),\ 
Z = \sum_{n=1}^N \exp \left( \triangle \beta S_n \right) 
~~{\rm and}~~ \triangle \beta = \beta - \beta_0 . \eqno(2.3) 
$$
For $\beta \ne \beta_0$ we call this reweighting.
For technical reasons (floating point number overflows) one is actually 
forced to use in practice $S_i' = S_i - {\overline S} (\beta_0 )$
in these formulas. This contributes an irrelevant multiplicative 
factor and makes the arithmetic computationally feasible. With
$f_n = \exp (i\beta_y S_n)$, the calculation of partition function 
zeros can be considered as a special case (where the numerator may
be omitted in the one histogram case). For biased quantities
the empirical error bar $\triangle {\overline f}$ of ${\overline f}$ 
is conveniently calculated by means of the jackknife method and
one may also correct for the remaining bias (see~[23] for details).

\midinsert
\bigskip
\centerline{\bf Table 1: Data and their reweighting ranges.}
$$
\vbox{\tabskip=0pt\offinterlineskip\tenpoint
   \halign{\strut\hfil#\hfil\tabskip=.7em&\vrule#&
     \hfil#\hfil&\vrule#&\hfil#\hfil&\vrule#&\hfil#\hfil&
       \vrule#&\hfil#\hfil&\vrule#&
      \hfil#\hfil&\vrule#&\hfil#\hfil&\vrule#&
          \hfil#\hfil\tabskip=0pt\cr
\tablerule
\strutrule
$L_t\cdot L^3$&&
\omit{\hfil therm.\hfil}&&
\omit{\hfil meas. (indep.)\hfil}&&
\omit{\hfil $\beta_0$\hfil}&&
\omit{\hfil $\beta_{\min}$\hfil}&&
\omit{\hfil $\beta_{\max}$\hfil}&&
\omit{\hfil $s_{q_1}$\hfil}&&
\omit{\hfil $s_{q_2}$\hfil}\cr
\tablerule
\omit\smallstrut&&&&&&&&&&&&&&\cr
\noalign{\vskip-3pt}
$~2\cdot ~6^3$ && 10,000 && 120,000 ~(360) && 5.094 && 5.026
                                         && 5.1550 && 0.4060 && 0.4710 \cr
$~2\cdot ~8^3$ && 10,000 && 120,000 ~(120) && 5.090 &&  5.042
                                         && 5.1220&&  0.4080 &&  0.4610 \cr
$~2\cdot 10^3$ && 10,000 && 120,000 ~~(30) && 5.090 &&  5.050
                                         &&  5.1080 &&  0.4090 &&  0.4550 \cr
$~2\cdot 12^3$ && 10,000 && 120,000 ~~(12) && 5.092 &&  5.064
                                         &&  5.1070 &&  0.4110 &&  0.4540 \cr
\deepstrut$~2\cdot 12^3$ && 10,000 && 120,000 ~~(12) && 5.095 &&  5.078
                                         &&  5.1140 &&  0.4130 &&  0.4580 \cr
\noalign{\vskip-2pt}
$~4\cdot ~4^3$ && 10,000 && 120,000 (1300) && 5.570 &&  5.488
                                         &&  5.6710 &&  0.4960  &&  0.5560 \cr
$~4\cdot ~4^3$ && 10,000 && 120,000 (2080) && 5.610 &&  5.533
                                         &&  5.7350 &&  0.5120  &&  0.5640 \cr
\deepstrut$~4\cdot ~4^3$ && 10,000 && 120,000 (3000) && 5.640 &&  5.566
                                         &&  5.7650 &&  0.5220  &&  0.5690 \cr
\noalign{\vskip-2pt}
$~4\cdot ~6^3$ && 10,000 && 120,000 (4000) && 5.500 &&  5.449
                                         &&  5.5510 &&  0.4830  &&  0.5110 \cr
$~4\cdot ~6^3$ && 10,000 && 120,000 (1200) && 5.640 &&  5.590
                                         &&  5.6880 &&  0.5230  &&  0.5520 \cr
$~4\cdot ~6^3$ && 10,000 && 120,000 (1400) && 5.645 &&  5.598
                                         &&  5.6970 &&  0.5240  &&  0.5540 \cr
$~4\cdot ~6^3$ && 10,000 && 120,000 (1400) && 5.660 &&  5.614
                                         &&  5.7170 &&  0.5290  &&  0.5570 \cr
$~4\cdot ~6^3$ && 10,000 && 120,000 (1600) && 5.690 &&  5.641
                                         &&  5.7540 &&  0.5380  &&  0.5640 \cr
\deepstrut$~4\cdot ~6^3$ && 10,000 && 120,000 (2800) && 5.740 &&  5.687
                                         &&  5.8090 &&  0.5500  &&  0.5720 \cr
\noalign{\vskip-2pt}
$~4\cdot ~8^3$ && 10,000 && 120,000 (2900) && 5.600 && 5.567
                                         && 5.6340 && 0.5160  && 0.5330 \cr
$~4\cdot ~8^3$ && 10,000 && 120,000 ~(800) && 5.670 && 5.638
                                         && 5.7030 && 0.5350  && 0.5540 \cr
$~4\cdot ~8^3$ && 10,000 && 120,000 ~(700) && 5.693 && 5.661
                                         && 5.7310 && 0.5420  && 0.5600 \cr
\deepstrut$~4\cdot ~8^3$ && 10,000 && 120,000 (1300) && 5.720 && 5.687
                                         && 5.7630 && 0.5500  && 0.5650 \cr
\noalign{\vskip-2pt}
$~4\cdot 10^3$ && 10,000 && 120,000 (3000) && 5.600 && 5.575
                                         && 5.6260 && 0.5180  && 0.5310 \cr
$~4\cdot 10^3$ && 10,000 && 120,000 ~(550) && 5.680 && 5.656
                                         && 5.7020 && 0.5400  && 0.5540 \cr
$~4\cdot 10^3$ && 10,000 && 120,000 ~(350) && 5.693 && 5.671
                                         && 5.7190 && 0.5440  && 0.5570 \cr
\deepstrut$~4\cdot 10^3$ && 10,000 && 120,000 ~(600) && 5.710 && 5.687
                                         && 5.7370 && 0.5490  && 0.5610 \cr
\noalign{\vskip-2pt}
$~4\cdot 12^3$ && 10,000 && 120,000 (3000) && 5.620 && 5.601
                                         && 5.6380 && 0.5250  && 0.5340 \cr
$~4\cdot 12^3$ && 10,000 && 120,000 ~(330) && 5.681 && 5.662
                                         && 5.6960 && 0.5410  && 0.5510 \cr
$~4\cdot 12^3$ && 10,000 && 120,000 ~(150) && 5.691 && 5.675
                                         && 5.7080 && 0.5440  && 0.5550 \cr
\deepstrut$~4\cdot 12^3$ && 10,000 && 120,000 ~(600) && 5.703 && 5.687
                                         && 5.7230 && 0.5490  && 0.5580 \cr
\noalign{\vskip-2pt}
$~4\cdot 14^3$ && 10,000 && 120,000 ~(240) && 5.682 && 5.668
                                         && 5.6930 && 0.5430  && 0.5510 \cr
$~4\cdot 14^3$ && 10,000 && 120,000 ~(110) && 5.691 && 5.678
                                         && 5.7030 && 0.5450  && 0.5540 \cr
\deepstrut$~4\cdot 14^3$ && 10,000 && 120,000 ~(440) && 5.698 && 5.687
                                         && 5.7130 && 0.5490  && 0.5560 \cr
\noalign{\vskip-2pt}
$~4\cdot 16^3$ && 15,000 && 120,000 ~(180) && 5.683  && 5.6711
                                         && 5.6923 && 0.5428 && 0.5499 \cr
$~4\cdot 16^3$ && 15,000 && 120,000 ~~(80) && 5.691  && 5.6793
                                         && 5.7006 && 0.5451 && 0.5536 \cr
$~4\cdot 16^3$ && 20,000 && 120,000 ~~(80) && 5.692  && 5.6814
                                         && 5.7034 && 0.5455 && 0.5541 \cr
\deepstrut$~4\cdot 16^3$ && 15,000 && 120,000 ~(200) && 5.697  && 5.6885
                                         && 5.7096 && 0.5485 && 0.5557 \cr
\noalign{\vskip-2pt}
$~4\cdot 20^3$ && 24,000 && 240,000 ~~(80) && 5.690  && 5.6820
                                         && 5.6959 && 0.5454 && 0.5522 \cr
$~4\cdot 20^3$ && 22,000 && 240,000 ~~(65) && 5.691  && 5.6833
                                         && 5.6975 && 0.5458 && 0.5528 \cr
\deepstrut$~4\cdot 20^3$ && 08,000 && 120,000 ~~(35) && 5.692  && 5.6842
                                         && 5.6990 && 0.5460 && 0.5529 \cr
\noalign{\vskip-2pt}
$~4\cdot 24^3$ && 30,000 && 180,000 ~~(44) && 5.691 && 5.6842 && 5.6945
                                                 && 0.5459 && 0.5512 \cr
$~4\cdot 24^3$ && 20,000 && 180,000 ~~(44) && 5.693 && 5.6890 && 5.6988
                                                 && 0.5477 && 0.5532 \cr
}}
$$
\bigskip
\endinsert

The relevant interval $\triangle\beta = \beta_{\max} - \beta_{\min}$
(reweighting range) will shrink with increasing volume 
like\footnote*{Note that $\sigma (s)$, the standard deviation of $s$,
goes as $L^{\rho/2}$, while
$ {d\beta \over d {\hat s}}|_{\beta=\beta_0} \sim L^{-\rho}$.}

$$ 
\triangle\beta\ \approx \sigma (s) 
\left. {d\beta \over d {\hat s}} \right|_{\beta=\beta_0}\
=\ ~{\rm const}\ L^{-\rho/2} L^{-d/2} 
=\ ~{\rm const}\ L^{-1/\nu} , \eqno(2.4) 
$$ 
where $\rho=\alpha/\nu$ and the last equality made use of the 
hyperscaling relation~[24] $\rho+d=2/\nu$.
We now have to determine the constant in~(2.4).
Essentially this boils down to the question: From which percentage
of our total data at $\beta_0$ do we still expect meaningful
results? Clearly, this depends somewhat on whether our statistics is
large or small. $q$-tiles $s_q$ of our empirical action density 
with $q$ in the range $0.025$ to $0.1$ may still be expected to give
meaningful answers. This has to be converted into a $\beta$-range. Let 
$q_1=q$ and $q_2=1-q$; we define $\beta_{\min}$ and $\beta_{\max}$ by:

$$ 
\beta_{\min} = \beta_{q_1} ~~~{\rm and}~~~ 
   \beta_{\max} = \beta_{q_2} ,                              
$$
where $\beta_q$ is given by the implicit equation

$$ 
{\hat s} (\beta_q)\ =\ s_q .                              
$$
This equation is solved numerically for $\beta_q$. Figures~$1-3$
illustrate this procedure for the $4\cdot 20^3$ lattice at 
$\beta_0 = 5.691$. Table~1 gives an overview of our SU(3) data 
including the $\beta_{\min}$, 
$\beta_{\max}$ and $s_{q_1}$, $s_{q_2}$ values for the choice $q=0.025$
(up to errors from conservative rounding of $\beta_{\min}$, 
$\beta_{\max}$).
For $L_t=4$ Figure~4 depicts the $\triangle\beta (L)$ ranges versus $L$.
Our subsequent analysis of autocorrelation times shows that for our 
present data the choice $q=0.025$ was too optimistic. However, this does
not really matter, because the main purpose of calculating 
$[\beta_{\min},\beta_{\max}]$ intervals first is to prevent reweighting
calculations at absurd $\beta$ values.

\vskip 4in
\figins{Action per plaquette: MC time series, $V=4\cdot
20^3$, $\beta_0 = 5.691$. Every 100th measurement is plotted here.}

\figins{Histogram with $q$-tiles $q=0.0285$ and
$q=0.0270$ for the action per plaquette data 
of figure~1. }

\figins{ Determination of the reweighting range.
Open squares denote the 
average action per plaquette 
${\overline s} (\beta )$, computed 
by reweighting from the MC simulation at the
$\beta_0$ corresponding to the full diamond.
The curve is an interpolation. The intersections
of this curve with the dashed lines corresponding
to the q-tiles determine 
$\beta_{\rm min}, \beta_{\rm max}$.}

\figins{ Reweighting ranges versus lattice size $L$ for all
our SU(3) data sets.}

In spin systems it is convenient to work with histograms. For lattice 
gauge theories the action varies typically over a continuous range and
a histogram method is not recommendable for two reasons:
\smallskip
\item{i)} The size of the histogram bin (i.e. of the action interval
          deemed to constitute a single histogram entry) is an
          extraneous parameter. It is tedious to have to cope with it.
\smallskip
\item{ii)} Whatever the size of the bin, inevitably part
              of the information contained in the original sample
              gets lost.
\smallskip
\noindent
Instead of artificially introducing histograms, it is more  
convenient to rely directly on the empirical time series for 
the data. This requires to keep all measurements on disk or tape.
In our present simulations we kept in double precision the spacelike and 
timelike plaquette expectation values and the real and imaginary Polyakov 
loop values. This amounts to up to $4*240,000$ Real*8 data per 
($L, \beta_0^i$) simulation point, altogether filling up about 0.2 
gigabyte of disk space (in unformatted storage). Consequently, the feasibility 
of this kind of analysis is tightly linked to the recent progress in 
storage technology and to the availability of large disks.

To cover a larger $\beta$ range one has to patch MC results from 
runs at different $\beta^i_0$ values ($\beta^{i+1}_0 > \beta^i_0$, 
$i=1,...,P$), whose validity ranges overlap:

$$ 
s^{i+1}_{q2} > s^{i}_{q_2} > s^{i+1}_{q_1} > s^{i}_{q1} . \eqno(2.5) 
$$
Various methods can be found in the literature~[3,4,8,9]. The two recent 
discussions~[8,9] both aim at minimizing the errors in the resultant 
estimates for $n(S)$. A crucial difference is that~[9] fixes the needed
relative normalizations of the histograms from data in the overlap
regions only, whereas~[8] exploits a self-consistency condition which
was previously stated in~[4]. The approach~[8] yields results even
if there is no overlap at all, whereas~[9] cannot be applied in such a
situation. For our purposes, results from patches without overlap are
assumed to be meaningless and applying the self-consistency condition
may be somewhat dangerous. For histograms with strong overlaps both methods
will converge towards identical results.

More generally, it is useful to patch in such a way 
that the error of the actually
calculated quantity is minimized. This leads to the following very
straightforward approach, which we present in a form suitable for the 
time series analysis of our data. 
The first observation is that any combination

$$ 
{\overline f} (\beta )
= {\sum_{i=1}^P a_i F_i \over \sum_{i=1}^P a_i Z_i}
~~~{\rm with\ weight\ factors}~~~ a_i = a_i(\beta ) > 0 \eqno(2.6) 
$$
is a valid estimator for $\langle f \rangle$. In the limit of infinite
statistics each single histogram would yield the correct results.
To simplify our considerations we impose the normalization

$$ 
\sum_{i=1}^P w_i = 1 ~~~{\rm with}~~~ w_i = a_i Z_i . \eqno(2.7) 
$$
This converts equation~(2.6) into

$$ 
{\overline f} = \sum_{i=1}^P w_i {\overline f}_i ~~~{\rm with}~~~
   {\overline f}_i = {F_i \over Z_i} . \eqno(2.8) 
$$
This equation makes clear that
 the optimal choice for the normalized weight
factors $w_i$ is simply the inverse variance of ${\overline f}_i$

$$ 
w_i\ \sim\ {1\over \sigma^2 ({\overline f}_i) } , \eqno(2.9a) 
$$
and the overall constant is fixed by the normalization condition~(2.7).
In practical applications the exact variances 
$\sigma^2 ({\overline f}_i)$ are unknown and we have to rely on the
empirical error bars as estimators:

$$ 
w_i\ \sim\ {1\over (\triangle {\overline f}_i )^2}~\ . \eqno(2.9b) 
$$
Of course, this just means that the standard way to
add estimators weighted by their error bars is also the most suitable 
one for 
combining estimators from MC simulations at various $\beta_0$ values.
However, several complications arise which deserve discussion.
 
\smallskip
\item{i)} Typically, our data exhibit large autocorrelation times
(see next section). This limits a serious statistical analysis to
using twenty jackknife
bins. Imposing a 95~\% confidence limit (more precisely 
the [0.025,0.975] confidence interval), the $\chi^2$ distribution~[25]
implies that $(\triangle {\overline f}_i)^2 /
\sigma^2 ({\overline f}_i)$ fluctuates in the range [0.58,2.13]. Our
experience is that the effect of these fluctuations on 
${\overline f} (\beta )$ is harmless as long as only data sets from
simulations at $\beta_0^i$ sufficiently close to $\beta$ are included. 
However, error bar fluctuations may become a serious problem if
equation (2.9b) is applied blindly. A highly recommendable cut-off is
to set $w_i=0$ for $\beta \notin [\beta_{\min},\beta_{\max}]$. It may
be suitable to constrain the included data sets even further. For 
instance by excluding data sets which:  a)~taken alone give unsatisfactory
results ${\overline f}_i (\beta )$ and b)~have $\beta$ located at one of
the edges of the validity range. 
 
\smallskip
\item{ii)} Once the weight factors are determined, error bars of
${\overline f} (\beta )$ from the combined statistics are not calculated
by the standard rules of error propagation. Instead, new bins are formed, 
each relying on the combined statistics~(2.6). If, after binning, 
autocorrelations still existed in the single data sets, they will 
become further reduced now as each new bin combines data from independent
simulations. When appropriate, the new bins are used to construct 
jackknife bins in the standard way.
 
\smallskip
\item{iii)} For connected estimators 
${\overline {f^c}} = {\overline {f^2}} - {\overline f}^2$, like the specific
heat or susceptibilities, it does not follow from equation (2.8) that
the optimal weight factor is $w_i \sim 1 / \sigma^2 ({\overline {f^c}}_i)$.
The reason is that one has to calculate ${\overline f}^2$ according
to ${\overline f}^2 = (\sum_i w_i {\overline f}_i)^2$ and not according
to ${\overline f}^2 = \sum_i w_i ({\overline f}_i)^2$. 
But patching ${\overline {f^c}}$ would require that ${\overline {f^2}}$
and ${\overline f}^2$ be calculated with {\it identical} 
weight factors.
Fortunately however, this
problem seems not to be too serious either. Weight factors calculated
from ${\overline f}_i, {\overline {f^2}}_i$ or ${\overline {f^c}}_i$ 
should differ little. We have already noted that there are substantial
statistical weight factor fluctuations which, in the reasonable range,
do not influence the final result significantly. Therefore, we decided
in favour of the simplest solution, namely to use the weight factors
$w_i \sim 1 / (\triangle {\overline f}_i )^2$ for the calculation of
${\overline {f^c}} (\beta)$.
\medskip 
\heading{AUTOCORRELATIONS IN COMPUTER TIME} 

It has been emphasized in recent literature~[18] that one has to control 
the integrated autocorrelation time $\hat{\tau}_{int}$, to be sure about
the final error bars. However, in a typical realistic situation 
a Monte Carlo simulation may be perfectly reasonable with respect 
to all calculated quantities, including the confidence intervals 
implied by their error bars. And yet, $\hat{\tau}_{int}$ may remain
the only quantity of interest that 
cannot be calculated reliably in such a simulation. It seems that 
previous investigations did not pay attention to this scenario and, 
therefore, we find it worthwhile to present some details.

To recall the concepts we first consider results obtained by a Metropolis 
generation of the Gaussian distribution. We generate $131,072 = 2^{17}$ 
numbers; due to a finite Metropolis stepsize, successive numbers are 
correlated. The integrated autocorrelation time is defined by:

$$ 
2\tau_{int}(n_b)\ =\ \rho (0) + 2\sum_{i=1}^{n_b} \rho (i) 
~~~{\rm with}~~~ \rho (i) = 
{\langle (s_0-{\hat s}) (s_i-{\hat s}) \rangle \over 
\langle  (s_0-{\hat s}) (s_0-{\hat s}) \rangle} . \eqno(3.1) 
$$
For $n_b\to\infty$ one recognizes that $2\tau_{int}$ is just the ratio 
of the correct variance $\sigma^2({\overline s})$ 
divided by the naive variance 
(obtained by using all events $s_n$ as if they were independent). 
A convenient way to calculate the correct variance is provided by
binning~[19]. The connection between binning and and the integrated
autocorrelation time has been emphasized in~[20]. Let us partition 
our data set into $N_b=2^{17-k}$ bins of length $n_b = 2^k$ ($N_b=32$
for $k=12$) and denote the variance of ${\overline s}$
after the $k^{th}$ binning by $\sigma^2_k({\overline s})$. Note that

$$
\sigma^2({\overline s})\ =\ 
\lim_{k\to\infty}\sigma^2_k ({\overline s}) , \eqno(3.2a) 
$$
whereas

$$ 
\sigma^2 (s)\ =\ \sigma^2_0 (s) .           \eqno(3.2b) 
$$
Now 

$$ 
{\sigma^2_k({\overline s}) \over \sigma^2_0({\overline s})}\ =\ 
\rho(0) + {2 (n_b-1) \over n_b} \rho (1) + {2 (n_b-2) \over n_b} 
\rho (2) + ... + {2 \over n_b} \rho (n_b-1) , \eqno(3.3) 
$$
and for $n_b\to\infty$ the approach towards $2\tau_{int}(n_b)$ follows 
from the rapid falloff of $\rho(i), i\to\infty$. We introduce the 
notation $(\triangle_k {\overline s})^2$ for the estimators corresponding 
to $\sigma^2_k ({\overline s})$.

\figins{ Metropolis generation of Gaussian random numbers: 
Integrated autocorrelation time by direct measurement 
$2\tau_{int} (n_b)$ (full circles) and by increase of 
the error bar under multiple binning 
$\triangle_k (\bar s) \over \triangle_0 (\bar s)$ (triangles)
versus $k$ (with $n_b=2^k$).}

Using the Gaussian Metropolis data, figure~5 compares the increase 
of the variance under multiple binning, 
$(\triangle_k \bar s)^2/(\triangle_0 \bar s)^2$,
with the direct calculation of the integrated autocorrelation time. As  
expected, for sufficiently large $n_b$ identical values ($\approx 3.7$) are 
approached. The convergence is somewhat better for the direct calculation 
of the integrated correlation time, whereas the binning procedure is 
computationally far more efficient.
As usual in this paper, the error bars of $2\tau_{int}$ are calculated 
with the double jackknife method~[23]. However, for the error bar of the 
empirical variance $(\triangle_k \bar s)^2$ we do not use an estimator, but 
assume instead the $\chi^2$ distribution with $N_b-1$ degrees of freedom 
for $(N_b-1) (\triangle_k \bar s)^2 / \sigma^2_k (\bar s)$. 
The error bars depicted in the
figures are then chosen such that twice their range, seen from the
center, corresponds to a normal 95.4~\% confidence interval. The 
depicted symbols are the centers of these error bars and not 
the actually calculated estimators 
$(\triangle_k \bar s)^2 / (\triangle_0 \bar s)^2$
(due to the asymmetry of the $\chi^2$ distribution, 
these have somewhat lower
values -- which quickly approach the center as $N_b\to\infty$).
If $n_b$ is
large, the bins do not only become statistically independent, but the 
central limit theorem implies at the same time that they become normally
distributed (each bin is an average over a large number of original data).
For the normal (Gaussian) distribution it is well known~[25] that the
variance is $\chi^2$ distributed. Therefore, our assumption is
justified whenever the statistical analysis is correct, i.e. sufficiently
large $n_b$ values can be reached such that the correlations between the 
bins can indeed be neglected.

\figins{ As in figure 5, for SU(3) MC data at $V=4\cdot 24^3$,
$\beta_0=5.691$. In the upper left corner an
enlargement of part of the figure is given.} 

Figure~6 is the analog of figure~5, calculated with our SU(3) action
data from the $4\cdot 24^3$ lattice at $\beta_0 = 5.691$. Obviously, 
satisfactory convergence is not achieved. It is remarkable that direct
calculation of the integrated autocorrelation time gives for large $n_b$
a much noisier estimator than we obtain from multiple binning. The
enlargement of the first part of figure~6 in its upper left corner 
demonstrates this most clearly. Presumably binned data are favourably
(de-)correlated. Consequently, we now rely on multiple
binning. For even larger $n_b$ (small number of bins $N_b$) also the 
error bars of $\sigma^2_k ({\overline s}) / \sigma^2_0 ({\overline s})$
increase rapidly. It is a perfectly possible
scenario that $N_b=20$ bins are already sufficiently independent to allow
a statistically reliable calculation of the action ${\overline s}$ (the 
$2\sigma$ confidence level of the Student distribution with $N_b=20$ is 
almost normal), but a self-consistent calculation of the integrated 
autocorrelation time is impossible. The binning procedure sheds light on
the reason. 
Let us assume we have $N_b=20$ independent bins. As already mentioned in
this paper, it follows from the $\chi^2$ distribution with $N_b-1$ degrees
of freedom that the 95~\% confidence interval of 
$(\triangle {\overline s})^2 / \sigma^2 ({\overline s})$ 
is [0.58,2.13]. In other words, the correct integrated autocorrelation 
time could be two times larger than the estimate or, equally
well, only half the estimate. (Remember that the error bars given in
the figures are half these confidence intervals.) To reduce this uncertainty
to approximately $\pm$15~\%, we need about $N_b=400$ independent bins. When,
like in gauge theories or especially in full QCD, the computational
effort is large, computer resources may only allow to create about $N_b=20$
independent bins. This may be fully sufficient to calculate quantities of
physical interest with reliable confidence intervals.  However, to compute
the integrated autocorrelation time reliably requires a factor 20 more 
computer time. We conclude that the demand to calculate the integrated 
autocorrelation time in each investigation is exaggerated.
Instead one may have to work under the assumption that the MC 
statistics is sufficiently large, for instance to give about
$N_b= 20$ independent bins, and then support this assumption with
a number of consistency checks. For example, one may perform
Student difference tests for estimators from independent MC runs
(see the subsequent sections). Such an assumption of a sufficiently
large statistics would be rather similar in spirit to other assumptions
(for instance about the approach towards the continuum limit)
already made for a typical lattice gauge theory simulation.

\figins{ As in figure 5, for SU(3) MC data at $V=4\cdot 6^3$,
$\beta_0=5.640$. }

As one expects, the situation is under better control for our smaller
systems. Figure~7 depicts the results from our $4\cdot 6^3$ lattice
at $\beta_0 =5.640$. While a direct calculation of the integrated
autocorrelation time remains inconclusive, its estimate from 
multiple binning is possible: $2\hat{\tau}_{int} = 100\pm 10$
is consistent with all results from $k=9$ ($N_b=234$) on.
Proceeding from smaller to larger lattices we obtain rough
estimates of $2\hat{\tau}_{int}$ for all our data sets. These results are
included in table~1 by defining the number of independent (indep.)
measurements as the number of measurements divided by $2\hat{\tau}_{int}$. 
One has to caution that in the case of $L=20$ (240,000 original measurements)
we can only rely on $N_b=14$ bins for estimating $2\hat{\tau}_{int}$.
This means an error of about $\pm$50~\% (around the center value) 
and no consistency checks towards higher $k$ values. 
The $L=24$ estimates may be unreliable
altogether. However, for our subsequent purposes the achieved overall
accuracy seems to be sufficient.

\medskip
\heading{SPECIFIC HEAT}
\midinsert 
\centerline{\bf Table 2: Results from the action.}
\medskip
\centerline{
\vbox{\tabskip=0pt\offinterlineskip\tenpoint
   \halign{\strut\hfil#\hfil\tabskip=.7em&\vrule#&
     \hfil#\hfil&\vrule#&\hfil#\hfil&\vrule#&\hfil#\hfil&
       \vrule#&\hfil#\hfil&\vrule#&
      \hfil#\hfil&\vrule#&\hfil#\hfil\tabskip=0pt\cr
\tablerule
\strutrule
$L_t\cdot L^3$&&
\omit{\hfil $\beta_0$\hfil}&& 
\omit{\hfil $\overline{ s}$\hfil}&&
\omit{\hfil $\sigma_0^2 (s)$\hfil}&&
\omit{\hfil $\sigma^2_{\max} (s) $\hfil}&&
\omit{\hfil bias \hfil}&&
\omit{\hfil $\beta_{\max}$\hfil}\cr
\tablerule
\omit\smallstrut&&&&&&&&&&&&\cr
\noalign{\vskip-3pt}
$~2\cdot ~6^3$ && 5.094 && 0.4390 (10) && 3.42\ $10^{-4}$ 
&& 3.44 (08) $10^{-4}$ && ~11 \% && 5.0917 (15) \cr
$~2\cdot ~8^3$ && 5.090 && 0.4329 (16) && 2.67\ $10^{-4}$ 
&& 2.70 (07) $10^{-4}$ && ~35 \% && 5.0908 (10) \cr
$~2\cdot 10^3$ && 5.090 && 0.4249 (26) && 2.05\ $10^{-4}$ 
&& 2.69 (11) $10^{-4}$ && 102 \% && 5.0928 (11) \cr
$~2\cdot 12^3$ && 5.092 && 0.4282 (31) && 2.31\ $10^{-4}$ 
&& 2.56 (07) $10^{-4}$ && 155 \% && 5.0928 (07) \cr
\deepstrut$~2\cdot 12^3$ && 5.095 && 0.4429 (25) && 1.61\ $10^{-4}$ 
&& 2.62 (31) $10^{-4}$ && ~62 \% && 5.0928 (10) \cr
\noalign{\vskip-2pt}
$~4\cdot ~4^3$ && 5.570 && 0.52781 (48) && 2.51\ $10^{-4}$ 
&& 2.698 (66)\ $10^{-4}$ && ~4 \% && 5.5446 (32) \cr
$~4\cdot ~4^3$ && 5.610 && 0.54101 (34) && 1.84\ $10^{-4}$ 
&& none         && &&       \cr
\deepstrut$~4\cdot ~4^3$ && 5.640 && 0.54822 (23) && 2.29\ $10^{-4}$ 
&& none         && &&       \cr
\noalign{\vskip-2pt}
$~4\cdot ~6^3$ && 5.500 && 0.49678 (13) && 5.32\ $10^{-5}$ 
&& none         && && \cr
$~4\cdot ~6^3$ && 5.640 && 0.53732 (25) && 5.91\ $10^{-5}$ 
&& 5.91 (12)\ $10^{-5}$ && ~6 \% && 5.638 (09) \cr
$~4\cdot ~6^3$ && 5.645 && 0.53905 (20) && 5.99\ $10^{-5}$ 
&& 6.17 (29)\ $10^{-5}$ && ~3 \% && 5.614 (14) \cr
$~4\cdot ~6^3$ && 5.660 && 0.54322 (18) && 5.62\ $10^{-5}$ 
&& none         && &&        \cr
$~4\cdot ~6^3$ && 5.690 && 0.55186 (14) && 4.66\ $10^{-5}$ 
&& none         && &&        \cr
\deepstrut$~4\cdot ~6^3$ && 5.740 && 0.56188 (12) && 3.47\ $10^{-5}$ 
&& none         && &&        \cr
\noalign{\vskip-2pt}
$~4\cdot ~8^3$ && 5.600 && 0.52471 (11) && 2.25\ $10^{-5}$
&& none         && &&        \cr
$~4\cdot ~8^3$ && 5.670 && 0.54429 (21) && 2.58\ $10^{-5}$ 
&& 2.59 (07)\ $10^{-5}$ && ~6 \% && 5.6713 (36) \cr
$~4\cdot ~8^3$ && 5.693 && 0.55098 (23) && 2.35\ $10^{-5}$ 
&& 2.49 (07)\ $10^{-5}$ && 10 \% && 5.6767 (52) \cr
\deepstrut$~4\cdot ~8^3$ && 5.720 && 0.55787 (10) && 1.65\ $10^{-5}$ 
&& none         && &&        \cr
\noalign{\vskip-2pt}
$~4\cdot 10^3$ && 5.600 && 0.52458 (07) && 1.11\ $10^{-5}$ 
&& none         && && \cr
$~4\cdot 10^3$ && 5.680 && 0.54664 (19) && 1.36\ $10^{-5}$
&& 1.406 (50)\ $10^{-5}$&& ~9 \% && 5.6876 (25) \cr
$~4\cdot 10^3$ && 5.693 && 0.55119 (17) && 1.28\ $10^{-5}$ 
&& 1.346 (50)\ $10^{-5}$&& ~6 \% && 5.6842 (32) \cr
\deepstrut$~4\cdot 10^3$ && 5.710 && 0.55558 (14) && 1.02\ $10^{-5}$ 
&& none         && &&        \cr
\noalign{\vskip-2pt}
$~4\cdot 12^3$ && 5.620 && 0.52998 (05) && 6.31\ $10^{-6}$ 
&& none         && && \cr
$~4\cdot 12^3$ && 5.681 && 0.54603 (18) && 7.85\ $10^{-6}$ 
&& 9.23 (55)\ $10^{-6}$ && 23~\% && 5.6909 (30) \cr
$~4\cdot 12^3$ && 5.691 && 0.54990 (27) && 9.35\ $10^{-6}$ 
&& 9.52 (33)\ $10^{-6}$ && 21~\% && 5.6884 (19) \cr
\deepstrut$~4\cdot 12^3$ && 5.703 && 0.55427 (12) && 6.42\ $10^{-6}$ 
&& none         && &&        \cr
\noalign{\vskip-2pt}
$~4\cdot 14^3$ && 5.682 && 0.54637 (19) && 5.40\ $10^{-6}$ 
&& 6.50 (32)\ $10^{-6}$ && 28 \% && 5.6882 (14) \cr
$~4\cdot 14^3$ && 5.691 && 0.54950 (30) && 6.99\ $10^{-6}$ 
&& 7.14 (36)\ $10^{-6}$ && 25 \% && 5.6924 (13) \cr
\deepstrut$~4\cdot 14^3$ && 5.698 && 0.55281 (13) && 4.51\ $10^{-6}$ 
&& none         && &&        \cr
\noalign{\vskip-2pt}
$~4\cdot 16^3$ && 5.683 && 0.54598 (14) && 3.21\ $10^{-6}$ 
&& 4.95 (42)\ $10^{-6}$ && 44 \% && 5.6902 (15) \cr
$~4\cdot 16^3$ && 5.691 && 0.54923 (28) && 5.16\ $10^{-6}$ 
&& 5.30 (26)\ $10^{-6}$ && 28 \% && 5.6921 (10) \cr
$~4\cdot 16^3$ && 5.692 && 0.54987 (28) && 5.40\ $10^{-6}$ 
&& 5.47 (26)\ $10^{-6}$ && 27 \% && 5.6916 (10) \cr
\deepstrut$~4\cdot 16^3$ && 5.697 && 0.55245 (13) && 3.47\ $10^{-6}$ 
&& none         && &&     \cr
\noalign{\vskip-2pt}
$~4\cdot 20^3$ && 5.690 && 0.54839 (22) && 3.32\ $10^{-6}$ 
&& 3.90 (16)\ $10^{-6}$ && 54 \% && 5.6918 (6) \cr
$~4\cdot 20^3$ && 5.691 && 0.54912 (28) && 3.88\ $10^{-6}$ 
&& 4.01 (15)\ $10^{-6}$ && 42 \% && 5.6915 (6) \cr
\deepstrut$~4\cdot 20^3$ && 5.692 && 0.54918 (32) && 3.64\ $10^{-6}$ 
&& 3.90 (24)\ $10^{-6}$ && 43 \% && 5.6929 (7) \cr
\noalign{\vskip-2pt}
$~4\cdot 24^3$ && 5.691 && 0.54781 (18) && 1.56\ $10^{-6}$
&& 4.4 (1.7)\ $10^{-6}$ && 46 \% && 5.6934 (9)\cr
$~4\cdot 24^3$ && 5.693 && 0.55101 (18) && 1.93\ $10^{-6}$ 
&& 2.89 (35)\ $10^{-6}$ && 46 \% && 5.6910 (8)\cr
}}}
\smallskip
\vtop{\tenpoint\noindent
The error bars are calculated with respect to twenty jackknife bins.
Estimators and error bars are corrected for the bias~[19]. 
For the maximum of the specific heat the
bias correction is given in percent of the error bar. The result ``none'' 
means that maximum of the specific heat is either unreliable or out of 
the controled $\beta$-range.}
\endinsert

To characterize phase transitions, action (energy) fluctuations are 
particularly suitable. The variance of the action
is related to the specific heat by 
$ c_V = \beta^{-2} \sigma^2 (S) /V $ .
The large $L$ finite size behaviour is 

$$ 
\sigma^2 (S)\ =\ \left< \sum_p \left( S_p - {\hat S_p} \right)
   \sum_p \left( S_p - {\hat S_p} \right) \right> 
$$

$$ 
=\ {\rm const}\ L^{\rho} \sum_p \langle (S_p - {\hat S_p})^2 \rangle\ 
   =\ {\rm const}\ L^{\rho}\ V_p \sigma^2 (S_p) 
~~~{\rm with}~~~ \rho = {\alpha\over\nu} ,  \eqno(4.1) 
$$
where $\sigma^2 (S_p)$ is a constant bounded by 
$\left( S_p^{\rm min} - S_p^{\rm max} \right)^2 /4$.
It is convenient to use the action density~(2.1):

$$ 
\sigma^2 (s)\ =\ V_p^{-2} \sigma^2 (S)\ =\ 
   {\rm const}\ L^{\rho}\ V_p^{-1} \sigma^2 (S_p) . \eqno(4.2) 
$$
The exponent $\rho$ differentiates three interesting classes:

$$ 
\cases{1.\ ~\rho = 0\ ~{\rm for\ non critical\ behaviour,}            \cr
       2.\ ~0 < \rho < d\ ~{\rm for\ second\ order\ critical\ behaviour,} \cr
          3.\ ~\rho = d\ ~{\rm for\ a\ first\ order\ phase\ transition.}}
    \eqno(4.3) 
$$ 
Here the given values of $\rho$ are defined as those obtained in the
limit $L\to\infty$. It is convenient to introduce the notion of a strong 
second order phase transition for transitions with $\rho$ less but close 
to $d$ (for instance $\rho=2.99$ and $d=3$). 
For second and first order transitions, the subleading correction to
equation~(4.3) is assumed to be non-critical. This leads to the 
following FSS fits

$$ 
\sigma^2 (s)\ =\ a_1 L^{\rho -d} + a_2 L^{-d} 
~~~{\rm for\ a\ second\ order\ phase\ transition}   \eqno(4.4a) 
$$
and

$$ 
\sigma^2 (s)\ =\ a_1 + a_2 L^{-d}
~~~{\rm for\ a\ first\ order\ phase\ transition} .  \eqno(4.4b) 
$$
The first order fit allows to determine the latent heat by the relation

$$ 
\triangle s\  =\ 2 \sqrt{a_1} .                  \eqno(4.5)  
$$
A weak first order transition is a first order transition with a small
latent heat. Numerically it is difficult to distinguish a weak first
order from a strong second order transition.
The FSS scaling relations~(4.4) are realized for $\beta = \beta_c$, 
where $\beta_c$ is the infinite volume critical coupling. 
In practice, the exact $\beta_c$ value is normally unknown.
But we can construct
the $L$ dependent series 
of $\beta_{\rm max} (L)$ values defined by

$$ 
\sigma^2 (s;L)\ (\beta )\ =\ {\rm maximum} ~~~~{\rm for}~~~
   \beta = \beta_{\rm max} (L)  .  \eqno(4.6) 
$$
The corresponding $\sigma^2 (s,\beta_{\max};L)$ is denoted by
$\sigma^2_{\max} (s)$. Eqs. (4.4) are also satisfied for this series.
In addition, this
approach yields a precise estimate of $\beta_c$ through the fit

$$ 
\beta_{\rm max} (L)\ =\ \beta_c + {\rm const}\ L^{-d} .\eqno(4.7) 
$$
Exponentially small corrections to this fit are of relevance for
small systems~[7], but are better neglected~[10] if the purpose is to 
estimate $\beta_c$. This is best done by fitting results from sufficiently 
large systems.

\midinsert 
\centerline{\bf Table 3:  Standard error propagation for the action
                          results.}
$$
\vbox{\tabskip=0pt\offinterlineskip\tenpoint
   \halign{\strut\hfil#\hfil\tabskip=.7em&\vrule#&
     \hfil#\hfil&\vrule#&\hfil#\hfil&\vrule#&\hfil#\hfil
          \tabskip=0pt\cr
\tablerule
\strutrule
$L_t\cdot L^3$&&
\omit{\hfil $\sigma^2_{\max} (s)$, goodness\hfil}&&
\omit{\hfil $\beta_{\max}$, goodness\hfil}&&
\omit{\hfil \# \hfil}\cr
\tablerule
\omit\smallstrut&&&&&&\cr
\noalign{\vskip-3pt}
$~2\cdot 12^3$&&2.56 (07)\ $10^{-4}$,~0.85&&5.0928 (06), ~0.94&& 2 \cr
&&&&&&\cr
$~4\cdot ~4^3$ &&2.70 (07)\ $10^{-4}$, ~$\,\;-\;\,$&&5.5446 (32), ~$\,-\;$
&& 1 \cr
$~4\cdot ~6^3$ && 5.95 (11)\ $10^{-5}$, ~0.41&& 5.631~ (08), ~0.16&& 2 \cr
$~4\cdot ~8^3$ && 2.54 (05)\ $10^{-5}$, ~0.32&& 5.6730 (30), ~0.40&& 2 \cr
$~4\cdot 10^3$ && 1.38 (04)\ $10^{-5}$, ~0.40&& 5.6863 (20), ~0.41&& 2 \cr
$~4\cdot 12^3$ && 9.44 (29)\ $10^{-6}$, ~0.65&& 5.6891 (16), ~0.49&& 2 \cr
$~4\cdot 14^3$ && 6.78 (24)\ $10^{-6}$, ~0.20&& 5.6905 (10), ~0.03&& 2 \cr
$~4\cdot 16^3$ && 5.32 (17)\ $10^{-6}$, ~0.57&& 5.6916 (07), ~0.57&& 3 \cr
$~4\cdot 20^3$ && 3.95 (10)\ $10^{-6}$, ~0.86&& 5.6920 (04), ~0.29&& 3 \cr
$~4\cdot 24^3$ && 2.95 (34)\ $10^{-6}$, ~0.40&& 5.6921 (06), ~0.05&& 2 \cr
}   
}   
$$
\vtop{\tenpoint\noindent
Results are calculated from table 2 according to standard error progagation.
The last column shows how many data sets were combined.}
\endinsert 

\midinsert 
\centerline{\bf Table 4:  Patching for the action results.}
$$
\vbox{\tabskip=0pt\offinterlineskip\tenpoint
   \halign{\strut\hfil#\hfil\tabskip=.7em&\vrule#&
     \hfil#\hfil&\vrule#&\hfil#\hfil&\vrule#&
#\hfil
          \tabskip=0pt\cr
\tablerule
\strutrule
$L_t\cdot L^3$&&
\omit{\hfil $\sigma^2_{\max} (s) $\hfil}&&
\omit{\hfil $\beta_{\max}$\hfil}&&
\omit{\hfil \# : weights. ~~~~~~\hfil}\cr
\tablerule
\omit\smallstrut&&&&&&\cr
\noalign{\vskip-3pt}
$~2\cdot 12^3$ && 2.55 (07)\ $10^{-4}$ && 5.0928 (07) && 2: 0.67, 0.33.\cr
&&&&&&\cr
$~4\cdot ~4^3$ && 2.71 (07)\ $10^{-4}$ && 5.5439 (34) && 2: 0.76, 0.24.\cr
$~4\cdot ~6^3$ && 6.03 (13)\ $10^{-5}$ && 5.624~ (11) && 2: 0.36, 0.64.\cr
$~4\cdot ~6^3$ && 6.03 (16)\ $10^{-5}$ && 5.620~ (11) && 3: 0.30, 0.53, 0.17.\cr
$~4\cdot ~8^3$ && 2.54 (06)\ $10^{-5}$ && 5.6719 (30) && 2: 0.61, 0.39.\cr
$~4\cdot 10^3$ && 1.37 (03)\ $10^{-5}$ && 5.6852 (18) && 2: 0.45, 0.55.\cr
$~4\cdot 12^3$ && 9.38 (29)\ $10^{-6}$ && 5.6896 (15) && 2: 0.34, 0.66.\cr
$~4\cdot 14^3$ && 6.70 (25)\ $10^{-6}$ && 5.6901 (09) && 2: 0.48, 0.52.\cr
$~4\cdot 14^3$ && 6.68 (20)\ $10^{-6}$ && 5.6895 (07) && 3: 0.39, 0.44, 0.17.\cr
$~4\cdot 16^3$ && 5.24 (15)\ $10^{-6}$ && 5.6916 (07) && 3: 0.17, 0.41, 0.43.\cr
$~4\cdot 16^3$ && 5.26 (14)\ $10^{-6}$ && 5.6913 (06) 
&& 4: 0.15, 0.35, 0.36, 0.14.\cr
$~4\cdot 20^3$ && 3.93 (09)\ $10^{-6}$ && 5.6920 (04) && 3: 0.38, 0.35, 0.27.\cr
$~4\cdot 24^3$ && 3.35 (17)\ $10^{-6}$ && 5.6919 (13) && 2: 0.42, 0.58.\cr
}   
}   
$$
\vtop{\tenpoint\noindent
The last column gives the number of patched data sets and the relative
weights, ordered by increasing $\beta_0$. When the numbers match, 
the combined data sets are identical with those processed in table~3.
Otherwise, the validity ranges of table~1 make it clear which
data sets have been added.
All error bars are calculated with respect to twenty jackknife bins
and corrected for the bias.} 
\endinsert 

For notational simplicity we drop henceforth
the distinction between estimators 
and exact theoretical definitions. Numerical results for each of our 
data sets are given in table~2. No stable values for $\sigma_{\max}$
are obtained when $\beta_0$ is too far from $\beta_{\max}$. Similarly,
insufficient statistics may cause unstable behavior. The $4\cdot 24^3$ 
lattice at $\beta_0 = 5.691$ is presumably an example. The discrepancies 
with~[17] are only due to the fact that we now use 20 jackknife bins,
whereas the error bars in~[17] were estimated with respect to 32
jackknife bins (on the large systems a too generous number). 
The results for $\sigma^2_{\max} (s)$ are calculated by means of 
the double jackknife approach~[23] and table 2 also lists the additional
correction for the bias in \% of the statistical error bar. Clearly,
statistical error and bias of the entire statistics have a similar order
of magnitude. For future reference we have also included the average
action ${\overline s} = {\overline s} (\beta_0 )$ and the variance
$\sigma_0^2 (s)$ at $\beta_0$ in table~2.

\midinsert 
\centerline{\bf Table 5: $L_{t}=4$ FSS fits of $\sigma^{2}_{\max}(s)$}
$$
\vbox{\tabskip=0pt\offinterlineskip\tenpoint
   \halign{\strut\hfil#\hfil\tabskip=.7em&
\vrule#&
\hfil#\hfil&
\hfil#\hfil&
\hfil#\hfil&
\vrule#&
\hfil#\hfil&
\hfil#\hfil&
\hfil#\hfil
\tabskip=0pt\cr
\tablerule
\strutrule
&&
\multispan3{\hfil $(Standard\;\;error\;\;propagation)$\hfil}&&
\multispan3{\hfil $(Patching)$\hfil}\cr
\noalign{\vskip-2pt}
$L$ range&&
\omit{\hfil $10^{6}a_{1}$\hfil}&
\omit{\hfil $10^{2}a_{2}$\hfil}&
\omit{\hfil $Q$\hfil}&&
\omit{\hfil $10^{6}a_{1}$\hfil}&
\omit{\hfil $10^{2}a_{2}$\hfil}&
\omit{\hfil $Q$\hfil}\cr
\tablerule
\omit\smallstrut&&&&&&&&\cr
\noalign{\vskip-3pt}
~4 - 24&&2.22(09)&1.26(02)&10$^{-23}$&&2.22(08)&1.27(2)&10$^{-25}$\cr
~6 - 24&&2.39(09)&1.20(02)&0.40&&2.42(08)&1.18(2)&0.25\cr
~8 - 24&&2.47(10)&1.17(03)&0.76&&2.49(09)&1.15(3)&0.81\cr
10 - 24&&2.48(12)&1.16(04)&0.64&&2.52(10)&1.14(3)&0.84\cr
12 - 24&&2.39(14)&1.22(06)&0.77&&2.46(12)&1.17(5)&0.91\cr
14 - 24&&2.41(18)&1.21(10)&0.58&&2.51(14)&1.14(8)&0.97\cr}}
$$
\vtop{
\tenpoint\noindent These fits use eq. (4.4b): thus, a first order 
transition is assumed.}
\endinsert

Using standard error propagation, we combine in table~3 for each lattice 
size the $\sigma^2_{\max} (s)$ and $\beta_{\max}$ estimates of table~2. 
Besides $\sigma^2_{\max} (s)$ and $\beta_{\max}$ the goodness of fit~[26]
is also listed. It is defined as the likelihood that the discrepancies 
between the estimates of table~2 (lattice size fixed) is due to chance.
In case of two data sets we use the standard Student test~[25] with
$N=20$ data. For more than two data sets we rely on $\chi^2$
and assume a Gaussian distribution. If the assumptions are correct (of
course there are slight corrections), the goodness of fit is a uniformly
distributed random number between zero and one. Except for the
goodness of fit for $\beta_{\max}$ from the $4\cdot 14^3$ and 
$4\cdot 24^3$ systems, all other values are reasonable. We tend to 
attribute the bad fit for the $4\cdot 14^3$ to statistical
fluctuations (after all we have a rather large number of systems), 
whereas a closer inspection (see below) of the $4\cdot 24^3$ data
sets gives rise to the suspicion that the assembled statistics is
insufficient in this case.

\figins{ Variance $\sigma^2 (s,\beta )$ from our single 
$4\cdot 24^3$ lattices. The full symbols indicate 
the $\beta_0$ values of the MC runs.}


\midinsert                                                          
\centerline{\bf Table 6: $L_{t}=4~\sigma^2(s)$ 
FSS fits of $\beta_{\max}(L)$}
$$
\vbox{\tabskip=0pt\offinterlineskip\tenpoint
   \halign{\strut\hfil#\hfil\tabskip=.7em&
\vrule#&
\hfil#\hfil&
\hfil#\hfil&
\hfil#\hfil&
\vrule#&
\hfil#\hfil&
\hfil#\hfil&
\hfil#\hfil
\tabskip=0pt\cr
\tablerule
\strutrule
&&
\multispan3{\hfil $(Standard\;\;error\;\;propagation)$\hfil}&&
\multispan3{\hfil $(Patching)$\hfil}\cr
\noalign{\vskip-2pt}
$L$ range&&
\omit{\hfil $\beta_c$\hfil}&
\omit{\hfil $a$\hfil}&
\omit{\hfil $Q$\hfil}&&
\omit{\hfil $\beta_c$\hfil}&
\omit{\hfil $a$\hfil}&
\omit{\hfil $Q$\hfil}\cr
\tablerule
\omit\smallstrut&&&&&&&&\cr
\noalign{\vskip-3pt}
~4 - 24&&5.6934(3)&-09.54(21.0)&0.20&&5.6933(3)&-09.60(22.0)&0.15\cr
~6 - 24&&5.6934(4)&-10.00(01.0)&0.15&&5.6936(4)&-10.60(01.1)&0.13\cr
~8 - 24&&5.6931(4)&-08.40(01.2)&0.60&&5.6933(4)&-09.30(01.2)&0.58\cr
10 - 24&&5.6928(5)&-06.10(01.8)&0.98&&5.6929(5)&-07.60(01.7)&0.80\cr
12 - 24&&5.6927(5)&-05.60(02.6)&0.96&&5.6929(6)&-07.50(02.5)&0.65\cr
14 - 24&&5.6926(6)&-04.90(03.4)&0.89&&5.6932(7)&-09.30(03.1)&0.66\cr}}
$$
\vtop{\tenpoint\noindent
The $\beta_c$ are extracted using eqs. (4.6) and (4.7).}
\endinsert 


\midinsert 
\centerline{\bf Table 7: $L_{t}=2$ FSS fits of $\sigma^{2}_{\max}(s)$ 
 and $\beta_{\max}(L)$}
$$
\vbox{\tabskip=0pt\offinterlineskip\tenpoint
   \halign{\strut\hfil#\hfil\tabskip=.7em&
\vrule#&
\hfil#\hfil&
\hfil#\hfil&
\hfil#\hfil&
\vrule#&
\hfil#\hfil&
\hfil#\hfil&
\hfil#\hfil
\tabskip=0pt\cr
\tablerule
\strutrule
&&
\multispan3{\hfil $(Standard\;\;error\;\;propagation)$\hfil}&&
\multispan3{\hfil $(Patching)$\hfil}\cr
\noalign{\vskip-2pt}
$L$ range&&
\omit{\hfil $10^{4}a_{1}$\hfil}& 
\omit{\hfil $10^{2}a_{2}$\hfil}& 
\omit{\hfil $Q$\hfil}&&      
\omit{\hfil $\beta_{c}$\hfil}&
\omit{\hfil $a$\hfil}&
\omit{\hfil $Q$\hfil}\cr
\tablerule
\omit\smallstrut&&&&&&\cr
\noalign{\vskip-3pt}
~6 - 12&&2.39(07)&2.17(25.0)&0.14&&5.0929(07)&-0.44(38.0)&0.36\cr
~8 - 12&&2.53(10)&0.94(71.0)&0.48&&5.0937(09)&-1.43(84.0)&0.61\cr
10 - 12&&2.38(23)&3.10(03.1)&-   &&5.0928(20)&~0.00(03.0)&-\cr
}}
$$
\vtop{\tenpoint\noindent
Fits to $\sigma^{2}_{\max}$ use eq. (4.4b); 
the $\beta_c$ are extracted using eqs. (4.6) and (4.7).}
\endinsert

Table 4 combines the action results by means of the patching method 
outlined in section~2. Within their statistical errors, the corresponding
estimates of tables~3 and~4 are fully compatible. Only for the 
$4\cdot 24^3$ lattices does patching reduce the error significantly.
Figure~8 depicts the $\sigma^2(s,\beta )$ reweighting calculation
for the single $4\cdot 24^3$ lattices, whereas figure~9 shows the 
patched result. From these figures it is clear that the real improvement
due to patching is in fact much more impressive than noticeable from
the $\sigma^2_{\max} (s)$ error bar reduction in table~4 as compared
with table~3 ~($4\cdot 24^3$ lattice). For our other data sets the reweighting 
analysis of single runs yields already reasonable $\sigma^2(s,\beta )$ 
pictures. Therefore, only one more illustration of patching: Figure~10 gives 
the combined $\sigma^2 (s,\beta )$ curve from all our four $4\cdot 16^3$ 
data sets.

\figins{ Variance $\sigma^2 (s,\beta )$ from our patched
$4\cdot 24^3$ lattices. The full symbols indicate 
the $\beta_0$ values of the MC runs.}

\figins{Variance $\sigma^2 (s,\beta )$ from our patched
$4\cdot 16^3$ lattices. The full symbols indicate 
the $\beta_0$ values of the MC runs.}

Let us first analyse the $L_t=4$ data.
Table~5 collects FSS fits~(4.4b) (which assume a first order transition)
for the $\sigma^2_{\max} (s)$ estimates
of tables~3 and~4. Obviously, results from lattices in the range from 
$L=8$ on are fully compatible with our hypothesis that the
transition is first order. If we now treat 
$\rho$ as a free parameter and perform second order fits~(4.4a), we get
$\rho = 3.08 \pm 0.23$ for the range $L=8-24$ (patched). Although
clearly consistent with $\rho = 3$, the accuracy is not very restrictive 
and the error becomes even worse if we restrict the fit to larger lattice 
sizes. Our best estimate of the latent heat is depicted in figure~11 and
yields

$$ 
\triangle s\  =\  (3.16\pm 0.06)\ 10^{-3}, ~~~(L_t=4).   \eqno(4.8) 
$$
Here we have used 
the $L = 8 - 24$ fit for the patched results, as given in table~5.
The patched results are preferred on the basis of
the arguments given in section~2. Monitoring the goodness of fit leads to 
the choice $L = 8 -24$. As soon as the goodness of fit is reasonable, there 
is no statistically significant inconsistency between the smaller and the
larger lattices. Still, omitting some smaller lattices in such a situation
may decrease the systematic errors. However, this is not relevant anymore
within the statistical accuracy. The main effect of omitting the smaller
lattices is only an increase of the statistical noise.

\figins{FSS fit eq. (4.4b) for the $L_t=4$ latent heat,
leading to the estimate~(4.8). }

Next, we use the $\beta_{\max} (L)$ estimates of tables~3
and~4 as input for the FSS fit~(4.7) and arrive at the results of table~6.
Our best estimate of $\beta_c$, corresponding to (4.8), is

$$ 
\beta_c^{\rm\ specific\ heat}\ 
=\ 5.6933 \pm 0.0004 , ~~~(L_t=4) . \eqno(4.9) 
$$
The line of arguments is similar as for our latent heat estimate.

The analysis of the $L_t=2$ data is straightforward. The first order
nature of the transition is much more pronounced than for $L_t=4$.
For the $2\cdot 10^3$ lattice the time series and corresponding action
histogram are depicted in figures~12 and ~13. This should be compared
with figures~1 and~2. In both cases we have the scale factor
$L/L_t = 5$, but only for $L_t = 2$ is the two-peak structure 
immediately clear. For the latent heat as well as for $\beta_c$ the
$L_t = 2$ FSS fits are now summarized in table~7. 
For $L=12$ (the only size for which we have two data sets), the results of
straightforward error propagation and patching are identical
(see tables~3 and~4). Thus we only need one table to show these fits.
Our best estimates (from $L=6-12$) are 

$$ 
\triangle s\  =\  (3.09 \pm 0.05)\ 10^{-2}, ~~~(L_t=2)   \eqno(4.10) 
$$
and

$$ 
\beta_c^{\rm\ specific\ heat}\ 
=\ 5.0929 \pm 0.0007 ,   ~~~(L_t=2) . \eqno(4.11) 
$$

 \figins{Action per plaquette: MC time series, 
$V=2\cdot 10^3$, $\beta_0 = 5.090$.
Every 50th measurement is plotted here.}
\figins{Histogram for the action per plaquette data 
of figure~12. }

\medskip
\heading{PARTITION FUNCTION ZEROS}

Ref.~[27] discusses the physical relevance of partition function zeros.
Their numerical calculation was pioneered in Refs. [5,28,29,21,9,17], 
of which [21,17] are concerned with SU(3) 
lattice gauge theory. In spin systems the action takes discrete values
and the partition function becomes a polynomial in $\exp (\beta )$. Under
such circumstances~[28,9] the Newton-Raphson method is convenient to
calculate the partition function zeros. When the action takes continuous 
values a time series analysis is more recommendable and we calculate 
the partition function zeros in two steps: first we scan graphically~[29]
for the separate zeros of the real and imaginary part. Figure~14 illustrates
this for our $4\cdot 20^3$ lattice at $\beta_0 = 5.692$. $Re (Z) = 0$ is
denoted by the crosses and $Im (Z)=0$ by the circles. A partition function
zero is obtained when the lines cross. Second, to compute the precise value 
for the leading zero, we then iterate with AMOEBA~[26] (with starting
values in a sufficiently small neighborhood of the zero).
 
\figins{ Partition function zeros for our $4\cdot 20^3$ lattice
at $\beta_0 = 5.692$. The circles indicate zeros for 
$Im (Z)$ and the crosses the zeros for $Re (Z)$. }

Before we can present our SU(3) results we have to clarify some subtle
details. For smaller SU(3) lattices we noted in~[30] that our empirical 
action distributions are well described by Gaussian fits. However, a 
Gaussian distribution does not give rise to zeros in the complex $\beta$ 
plane. Nevertheless we have reported zeros in~[17]. To resolve this 
paradoxical situation, we first study Gaussian random numbers and proceed 
then with the analysis of our SU(3) action distributions.

\midinsert 
\centerline{\bf Table 8: Partition function zeros.}
\medskip
\centerline
{\vbox{\tabskip=0pt\offinterlineskip\tenpoint
   \halign{\strut\hfil#\hfil\tabskip=.7em&\vrule#&
     \hfil#\hfil&\vrule#&\hfil#\hfil&\vrule#&\hfil#\hfil&
       \vrule#&\hfil#\hfil&\vrule#&
      \hfil#\hfil&\vrule#&\hfil#\hfil&\vrule#&\hfil#\hfil\tabskip=0pt\cr
\tablerule
\strutrule
$L_t\cdot L^3$&&
\omit{\hfil $\beta_0$\hfil}&& 
\omit{\hfil $\beta^0_x$\hfil}&&
\omit{\hfil $\beta^0_y$\hfil}&&
\omit{\hfil ${\rm bias}_x$\hfil}&&
\omit{\hfil ${\rm bias}_y$\hfil}&&
\omit{\hfil $\triangle\beta^{\max}_x$\hfil}&&
\omit{\hfil $\beta^{\max}_y$\hfil}\cr
\tablerule
\omit\smallstrut&&&&&&&&&&&&&&\cr
\noalign{\vskip-5pt}
$~2\cdot ~6^3$ && 5.0940 && 5.0910 (18) && 0.03960 (10) 
&& ~~~3 \% && ~~-6 \% && 0.03150 && 0.05050 \cr
$~2\cdot ~8^3$ && 5.0900 && 5.0905 (11) && 0.01759 (36) 
&& ~~~2 \% && ~-10 \% && 0.01290 && 0.02180 \cr
$~2\cdot 10^3$ && 5.0900 && 5.0927 (11) && 0.00865 (14) 
&& ~~-8 \% && ~~-6 \% && 0.00640 && 0.01040 \cr
$~2\cdot 12^3$ && 5.0920 && 5.0928 (08) && 0.00502 (08) 
&& ~~-4 \% && ~~-9 \% && 0.00080 && 0.00500\cr
\deepstrut$~2\cdot 12^3$ && 5.0950 && 5.0929 (11) && 0.00495 (11) 
&& ~~23 \% && ~-28 \% && 0.00350 && 0.00570 \cr
\noalign{\vskip-2pt}
$~4\cdot ~4^3$ && 5.5700 && 5.5500 (03) && 0.09800 (04)
&& ~~-2 \% && ~~-5 \% && 0.05000 && 0.10800 \cr
$~4\cdot ~4^3$ && 5.6100 && 5.5600 (05) && 0.14900 (07) 
&& ~-17 \% && ~~~2 \% && 0.0000$^*$ && 0.1230$^*$ \cr
\deepstrut$~4\cdot ~4^3$ && 5.6400 && 5.6070 (06) && 0.18500 (06) 
&& ~~-7 \% && ~~~7 \% && 0.0000$^*$ && 0.1170$^*$ \cr
\noalign{\vskip-2pt}
$~4\cdot ~6^3$ && 5.5000 && none&&&&&& &&&& \cr
$~4\cdot ~6^3$ && 5.6400 && 5.6540 (10) && 0.05200 (25)
&& ~104 \% && -140 \% && 0.04200 && 0.06500  \cr
$~4\cdot ~6^3$ && 5.6450 && 5.6560 (05) && 0.07570 (64)
&& ~-50 \% && ~-22 \% && 0.0000$^*$ && 0.0660$^*$ \cr
$~4\cdot ~6^3$ && 5.6600 && 5.6420 (07) && 0.07840 (47)
&& ~~10 \% && ~~35 \% && 0.0000$^*$ && 0.0670$^*$ \cr
$~4\cdot ~6^3$ && 5.6900 && 5.6450 (08) && 0.08010 (80)
\deepstrut&& ~-12 \% && ~-11 \% && 0.0000$^*$ && 0.0620$^*$ \cr
\noalign{\vskip-2pt}
$~4\cdot ~6^3$ && 5.7400 && none&&&&&& &&&& \cr
$~4\cdot ~8^3$ && 5.6000 && none&&&&&& &&&& \cr
$~4\cdot ~8^3$ && 5.6700 && 5.6747 (23)&& 0.04660 (27)
&& ~~-2 \% && ~-12 \% && 0.0000$^*$ && 0.0410$^*$ \cr
$~4\cdot ~8^3$ && 5.6930 && 5.6791 (33)&& 0.04980 (42)
\deepstrut&& ~~~5 \% && ~-16 \% && 0.0000$^*$ && 0.0420$^*$ \cr
\noalign{\vskip-2pt}
$~4\cdot 8^3$ && 5.7200 && none&&&&&& &&&& \cr
$~4\cdot 10^3$ && 5.6000 && none&&&&&& &&&& \cr
$~4\cdot 10^3$ && 5.6800 && 5.6889 (14)&& 0.03010 (18)
&& ~~~8 \% && ~-11 \% && 0.0000$^*$ && 0.0270$^*$ \cr
$~4\cdot 10^3$ && 5.6930 && 5.6864 (55)&& 0.02800 (81)
&& ~~~7 \% && ~~-7 \% && 0.0030$^*$ && 0.0270$^*$ \cr
\deepstrut$~4\cdot 10^3$ && 5.7100 && none &&&&      &&   && && \cr
\noalign{\vskip-2pt}
$~4\cdot 12^3$ && 5.6200 && none&&&&&& &&&& \cr
$~4\cdot 12^3$ && 5.6810 && none&&&&&&&&&& \cr
$~4\cdot 12^3$ && 5.6910 && 5.6896 (17)&& 0.02030 (07)
&& ~~~6 \% && ~~-3 \% && 0.0000$^*$ && 0.0180$^*$ \cr
\deepstrut$~4\cdot 12^3$ && 5.7030 && none&&&&&&&&&& \cr
\noalign{\vskip-2pt}
$~4\cdot 14^3$ && 5.6820 && 5.6886 (18)&& 0.01430 (09)
&& ~~-2 \% && ~-13 \% && 0.0050$^*$ && 0.0140$^*$ \cr
$~4\cdot 14^3$ && 5.6910 && 5.6922 (13)&& 0.01380 (07)
&& ~~~0 \% && ~-13 \% && 0.0000$^*$ && 0.0120$^*$ \cr
\deepstrut$~4\cdot 14^3$ && 5.6980 && 5.6859 (23)&& 0.02020 (28)
&& ~-69 \% && ~~87 \% && 0.0000$^*$ && 0.0130$^*$ \cr
\noalign{\vskip-2pt}
$~4\cdot 16^3$ && 5.6830 && 5.6904 (16)&& 0.01010 (10)
&& ~-11 \% && ~-15 \% && 0.00810 && 0.01060 \cr
$~4\cdot 16^3$ && 5.6910 && 5.6918 (10)&& 0.01010 (06)
&& ~~-2 \% && ~-13 \% && 0.0000$^*$ && 0.0094$^*$ \cr
$~4\cdot 16^3$ && 5.6920 && 5.6917 (10) && 0.00960 (05)
&& ~~-2 \% && ~-14 \% && 0.0000$^*$ && 0.0092$^*$ \cr
\deepstrut$~4\cdot 16^3$ && 5.6970 && none&&&&&& &&&& \cr
\noalign{\vskip-2pt}
$~4\cdot 20^3$ && 5.6900 && 5.6917 (06)&& 0.00554 (22)
&& ~~-5 \% && ~-16 \% && 0.00230 && 0.00570 \cr
$~4\cdot 20^3$ && 5.6910 && 5.6915 (06)&& 0.00527 (17)
&& ~~~1 \% && ~~-6 \% && 0.00120 && 0.00540 \cr
\deepstrut$~4\cdot 20^3$ && 5.6920 && 5.6929 (07)&& 0.00531 (28)
&& ~~-1 \% && ~-15 \% && 0.0000$^*$ && 0.0051$^*$ \cr
\noalign{\vskip-2pt}
$~4\cdot 24^3$ && 5.6910 && 5.6931 (07)&& 0.00270 (02)
&& ~-29 \% && ~-35 \% && 0.00390 && 0.00420 \cr
$~4\cdot 24^3$ && 5.6930 && 5.6913 (09)&& 0.00320 (04)
&& ~~32 \% && ~-49 \% && 0.00280 && 0.00390 \cr
}}}
\smallskip
\vtop{\tenpoint\noindent
The error bars are calculated with respect to twenty jackknife bins.
The $4\cdot 4^3$ estimates of this table are different from~[17],
where we overestimated the confidence radius and conjectured
a non-leading zero
to be the correct continuation of the results from
the larger lattices. Otherwise the differences with~[17] are small and
entirely due to using a different number of jackknife bins.}
\endinsert 

\subheading{The Gaussian Distribution}

Assume a lattice gauge theory MC simulation ($V_p =$ number of plaquettes)
and let 

$$ 
x\ =\ s- {\hat s} . \eqno(5.1) 
$$
For non-critical behaviour the measured probability density for $x$ 
will be 

$$ 
P(x)\ =\ \sqrt{A\over\pi} \exp \left( - A x^2 \right) 
~~~{\rm with}~~~ A = {1\over 2 \sigma^2} = a V_p . \eqno(5.2) 
$$
By reweighting with $\beta = \beta_x + i\beta_y$ we obtain the
partition function up to a normalization factor:

$$ 
z(\beta )\ =\ {Z(\beta )\over Z(\beta_0)}\ =
   \sqrt{A\over\pi} \int_{-\infty}^{+\infty} \exp (Bx+iCx)\
   \exp(-Ax^2)\ dx                                    \eqno(5.3a) 
$$
with (defining $b_x$ and $b_y$)

$$ 
B = V_p (\beta_x - \beta_0) =: V_p b_x
~~~{\rm and}~~~  C = V_p \beta_y =: V_p b_y .     \eqno(5.3b) 
$$
Integration over $x$ gives

$$ 
z(\beta )\ =\ \exp \left({B^2-C^2\over 4A}\right) 
\exp \left( i{BC\over 2A } \right)\ =\  |z(\beta )| 
\left[ \cos \left({BC\over 2A}\right) 
+ i \sin \left({BC\over 2A}\right) \right] . \eqno(5.4) 
$$
We have zeros of the imaginary part for

$$ 
B = 0 ~~~{\rm and}~~~ C = {2n\pi A \over B},\ (n=1,2,...) \eqno(5.5) 
$$
and of the real part for

$$ 
C = {(2n+1)\pi A \over B},\ (n=0,1,2,...) . \eqno(5.6) 
$$
Rewriting equations (5.5-5.6) in terms of $b_x, b_y$ we obtain
zeros of $Im (Z)$ for

$$ 
b_x = 0 ~(b_y ~{\rm arbitrary}) ~~~{\rm and}~~~ 
   b_y = {2n\pi A \over (V_p)^2 b_x} 
       = {2n\pi a \over V_p b_x},\ (n=1,2,...)   \eqno(5.5') 
$$
and of $Re (Z)$ for

$$ b_y = {(2n+1)\pi A \over (V_p)^2 b_x}
= {(2n+1)\pi a \over V_p b_x} ,\ (n=0,1,2,...) . \eqno(5.6') 
$$

The variance $\sigma^2(|z(\beta )|)$ is easily calculated to be

$$ 
\sigma^2 \left( |z(\beta )| \right)\ =\ 
\exp \left( {B^2 \over A}\right) - |z(\beta )|^2 . \eqno(5.7) 
$$
Assume that a numerical calculation has generated $N$ independent data
with the probability density~(5.2). We trust with an about 84\% 
(one sided!) confidence level 
that the calculation of $z(\beta )$ will not produce artificial
zeros in the $(B,C)$-range determined by

$$ 
\sigma \left( | \overline{z} (\beta ) | \right)\ =\
   {\sigma \left( |z (\beta )| \right) \over \sqrt{N}}\ \le\
   |z(\beta )| .                                   \eqno(5.8) 
$$
Defining

$$ 
X\ =\ \exp \left( {B^2 \over 4A} \right) ~~~{\rm and}~~~
   Y\ =\ \exp \left({-C^2 \over 4A} \right) , 
$$
the equality of (5.8) becomes

$$ 
X\ =\ \left( N +1 \right)^{1/2}\ Y, ~~~~{\rm where}~~~ 
\left( N+1 \right)^{-1/2}\ \le\ Y\ \le\ 1.         \eqno(5.9) 
$$  
The argument is only approximate, since numerical results within this
confidence radius may have some degree of independence, which is
difficult to assess. Here they are just treated as one event.

To give a numerical example, we take $A=1/(2\sigma^2) = 80,000$ 
and $V_p=6\cdot 4\cdot 14^3 = 65,856$.
We use a Gaussian pseudo random number generator and generate MC
data according to the probability density~(5.2). Subsequently, a
reweighting analysis is done to determine the zeros. For 1,200
and 76,800 independent data, respectively, figures~15
and~16 compare the exact values for the leading zeros of $Im (Z)$
and $Re (Z)$ with those from the simulation. Using equation~(5.9)
the expected range of validity for the numerical results is
also indicated and found to be respected by the data.
Namely, the apparent crossings (zeros of the partition function)
are seen to fall outside the confidence radius. In the Gaussian case
we know for sure that this means they are numerical fakes. 
For our $SU(3)$ data, we shall therefore have to reject any zeros
which fall outside the confidence radius.
 
\figins{ Simulation with 1,200 data. The circles indicate zeros
for $Im (Z)$ and the crosses zeros
for $Re (Z)$. The five ``longitudinal'' curves
depict corresponding exact results. The two
``latitudinal''
curves denote the validity range~(5.9) with $N=1,200$
(lower curve) and $N=76,800$ (upper curve).}
\figins{ As in figure~15, but relying on 76,800 data, and the
validity range for 1,200 data is no longer given.}

\midinsert 
\centerline{\bf Table 9: Patching of Partition Function Zeros.}
$$
\vbox{\tabskip=0pt\offinterlineskip\tenpoint
   \halign{\strut\hfil#\hfil\tabskip=.7em&
\vrule#&
     \hfil#\hfil&
\vrule#&
   \hfil#\hfil&
\vrule#&
\hfil#\hfil&
\vrule#&
   \hfil#\hfil&
\vrule#&
#\hfil
          \tabskip=0pt\cr
\tablerule
\strutrule
$L_t\cdot L^3$&&
\omit{\hfil $\beta^0_x$\hfil}&&
\omit{\hfil $\beta^0_y$\hfil}&&
\omit{\hfil [$\beta^{\min}_x,\beta^{\max}_x$] \hfil}&&
\omit{\hfil $\beta^{\max}_y$\hfil}&&
\omit{\hfil \# : weights.~~~~~~~~~~~~~~~~\hfil}\cr
\tablerule
\omit\smallstrut&&&&&&&&&\cr
\noalign{\vskip-3pt}
$~2\cdot ~12^3$ && 5.0928 && 0.00501 (7)
&& [5.090,5.095] && 0.0060 && 2: 0.63, 0.37. \cr
&& && && && && \cr
$~4\cdot ~4^3$ && 5.5520 (3) && 0.12300 (6)
&& none && 0.117$^*$ && 3: 0.74, 0.24, 0.02. \cr
$~4\cdot ~6^3$ && 5.6500 (7) && 0.07800 (5)
&& none && 0.073$^*$ && 4: 0.29, 0.37, 0.32, 0.02. \cr
$~4\cdot ~8^3$ && 5.6740 (2) && 0.04500 (2)
&& none && 0.042$^*$ && 2: 0.68, 0.32. \cr
$~4\cdot 10^3$ && unstable   && unstable 
&& none && none && 2: 0.52, 0.48. \cr
$~4\cdot 14^3$ && 5.6890 (2) && 0.01450 (6)
&& [5.687,5.691] && 0.0156 && 3: 0.26, 0.44, 0.30. \cr
$~4\cdot 16^3$ && 5.6915 (7) && 0.00990 (4)
&& [5.688,5.693] && 0.0106 && 3: 0.09, 0.45, 0.46. \cr
$~4\cdot 20^3$ && 5.6921 (5) && 0.00550 (2) 
&& [5.690,5.694] && 0.0058 && 3: 0.27, 0.42, 0.31. \cr
$~4\cdot 24^3$ && 5.6928 (6) && 0.00290 (2)
&& [5.689,5.695] && 0.0044 && 2: 0.45, 0.55. \cr
}   
}   
$$
\vtop{\tenpoint\noindent
We patch those data sets for which results are also reported in
table~8. The weights are in order of increasing $\beta_0$. Instead
of $\triangle \beta^{\max}_x$ we report now $\left[ \beta_x^{\min},
\beta_x^{\max} \right]$ as $\beta_0$ is no longer unique.}
\endinsert

\subheading{SU(3) results}

For single runs our leading partition function zeros are collected 
in table~8. To estimate whether they are inside or outside the 
confidence radii defined by equation~(5.9), we use the estimated
number of independent measurements from table~1 (for instance 44 for
the $4\cdot 24^3$ lattices) and $\sigma^2_0 (s)$ from table~2 as
width of a fictitious Gaussian distribution. This leads to the
$\triangle \beta_x^{\max}$ and $\beta_y^{\max}$ values
reported in table~8. An example of a zero and its confidence radius
is given in figure~17 (also showing the definition of 
$\triangle \beta_x^{\max}$ and $\beta_y^{\max}$).
An asterix in table~8 indicates that the 
estimated zero lies actually outside the radius of confidence. We
see that for the $4\cdot L^3$ lattices most results have problems. The
issue is actually quite subtle as repetitions with similar statistics 
lead to reproducible results. The reason is that for $\beta_x = \beta_0$
and a Gaussian distribution, $Z (\beta_y)$ falls off exponentially with
increasing $\beta_y$. As soon as the statistical noise becomes large
enough this leads with certainty to a fake crossover of real and imaginary 
zeros, as illustrated in figures~14 and~16. Upon a closer inspection of
table~8 one may argue that the $2\cdot 12^3$, $\beta_0 = 5.092$ data
set has also a problem. However, for the $L_t=2$ distributions we have
a pronounced double peak structure and the use of $\sigma_0 (s)$ from
table~2 is not justified. Each of the single Gaussians has a much smaller
width, leading to confidence radii larger than those reported in 
table~8.

\figins{ The leading zero and its confidence radius for
the $4\cdot 24^3$ lattice at $\beta_0=5.691$.
In the figure $\beta_0$ is scaled to zero and the
definition of $\triangle \beta^{\max}_x$ and
$\beta^{\max}_y$ is outlined.}

To rescue some of our estimates for the $4\cdot L^3$ lattices, we appeal
to our patching method. The confidence radii for the patched results
are estimated by iterating the equation

$$ 
\sqrt{ \sum_{i=1}^P (w_i)^2 \sigma_i^2 (| {\overline z} (\beta) | )}\
=\ \sum_{i=1}^P w_i |z_i (\beta )| . \eqno(5.10) 
$$  
The final results are collected in table~9. Whereas for lattices of size
$4\cdot 4^3$ to $4\cdot 12^3$ we still have no conclusive results, we
can now determine the leading zero on lattices $4\cdot L^3$ with 
$L\ge 14$. For these lattices the FSS fit

$$ 
u^0_y (L)\ \sim\ L^{-1/\nu} ~~~{\rm with}~~~ u\ =\ e^{-\beta} \eqno(5.11) 
$$
gives

$$ 
\nu\ =\ 0.35 \pm 0.02,\ (L_t=4)  \eqno(5.12) 
$$
with a goodness of fit $Q=0.26$. This is consistent with $\nu=1/d$, 
i.e. with a first order transition.
Fitting $\beta_x^0 (L)$ with the
FFS formula~(4.7) yields another estimate of the infinite volume critical
point:

$$ 
\beta_c^{\rm\ zeros}\ =\ 5.6934 \pm 0.0007,\ (L_t=4). \eqno(5.13) 
$$

\noindent{The fitted range is $L=14-24$ and the 
goodness of fit is $Q=0.73$.}

Compared with the $L_t=4$ situation, the $L_t=2$ case is unproblematic.
All our zeros from $L=6$ to $L=12$ are within the radii of convergence 
and allow the FSS fit~(5.11)

$$ 
\nu\ =\ 0.332 \pm 0.004,\ (L_t=2) \eqno(5.14) 
$$
with the acceptable goodness of fit $Q=0.12$. Fitting only the range
$L=8-12$ gives $\nu = 0.323 (6)$ with the goodness of fit $Q=0.30$.
The result~(5.14) confirms (with less effort)
the pioneering work of Karliner et 
al.~[21], who reported $\nu = 0.331 (6)$ from their more complicated
constrained MC calculation of partition function zeros. 
Our $L=6-12$ estimate of the critical $\beta$ is

$$ 
\beta_c^{\rm\ zeros}\ =\ 5.0930 \pm 0.0007,\ (L_t=2). \eqno(5.15) 
$$
with a goodness of fit $Q=0.42$. We see that the $\beta_c^{\rm\ zeros}$
estimates are well consistent with the $\beta_c^{\rm\ specific\ heat}$
results of section~4.


\midinsert 
\centerline{\bf Table 10: Results from the Polyakov susceptibility.}
\smallskip
\centerline{
\vbox{\tabskip=0pt\offinterlineskip\tenpoint
   \halign{\strut\hfil#\hfil\tabskip=.7em&\vrule#&
     \hfil#\hfil&\vrule#&\hfil#\hfil&\vrule#&\hfil#\hfil&
       \vrule#&\hfil#\hfil&\vrule#&
      \hfil#\hfil&\vrule#&\hfil#\hfil\tabskip=0pt\cr
\tablerule
\strutrule
$L_t\cdot L^3$&&
\omit{\hfil $\beta_0$\hfil}&& 
\omit{\hfil $\overline{ P}$\hfil}&&
\omit{\hfil $L^{-3}\chi_0(P)$\hfil}&&
\omit{\hfil $L^{-3}\chi_{\max}(P) $\hfil}&&
\omit{\hfil bias \hfil}&&
\omit{\hfil $\beta_{\max}$\hfil}\cr
\tablerule
\omit\smallstrut&&&&&&\cr
\noalign{\vskip-6pt}
$~2\cdot ~6^3$ && 5.094 && 0.77480 (0211) && 1.44\ $10^{-1}$ 
&& 1.52 (003) $10^{-1}$ && ~17 \% && 5.0836 (25) \cr
$~2\cdot ~8^3$ && 5.090 && 0.62510 (0378) && 1.58\ $10^{-1}$ 
&& 1.60 (004) $10^{-1}$ && ~38 \% && 5.0894 (12) \cr
$~2\cdot 10^3$ && 5.090 && 0.38700 (0700) && 1.49\ $10^{-1}$ 
&& 1.87 (006) $10^{-1}$ && 122 \% && 5.0936 (16) \cr
$~2\cdot 12^3$ && 5.092 && 0.84400 (0680) && 1.13\ $10^{-1}$ 
&& 1.95 (005) $10^{-1}$ && 177 \% && 5.0928 (07) \cr
\deepstrut$~2\cdot 12^3$ && 5.095 && 0.45420 (0872) && 1.79\ $10^{-1}$ 
&& 2.01 (028) $10^{-1}$ && ~53 \% && 5.0922 (12) \cr
\noalign{\vskip-3.5pt}
$~4\cdot ~4^3$ && 5.570 && 0.27737 (0296) && 2.67\ $10^{-2}$ 
&& none         && &&       \cr
$~4\cdot ~4^3$ && 5.610 && 0.33716 (0404) && 3.34\ $10^{-2}$ 
&& none         && &&       \cr
\deepstrut$~4\cdot ~4^3$ && 5.640 && 0.35873 (0370) && 3.53\ $10^{-2}$ 
&& none         && &&       \cr
\noalign{\vskip-3.5pt}
$~4\cdot ~6^3$ && 5.500 && 0.09542 (0050) && 0.26\ $10^{-2}$ 
&& none         && && \cr
$~4\cdot ~6^3$ && 5.640 && 0.21732 (0437) && 1.81\ $10^{-2}$ 
&& none         && && \cr
$~4\cdot ~6^3$ && 5.645 && 0.23119 (0455) && 1.99\ $10^{-2}$ 
&& 3.06 (014)\ $10^{-2}$ && ~8 \% && 5.6950 (06) \cr
$~4\cdot ~6^3$ && 5.660 && 0.27107 (0430) && 2.38\ $10^{-2}$ 
&& 2.90 (007)\ $10^{-2}$ && ~5 \% && 5.6970 (04) \cr
$~4\cdot ~6^3$ && 5.690 && 0.37507 (0670) && 2.87\ $10^{-2}$ 
&& 2.88 (007)\ $10^{-2}$ && ~6 \% && 5.6960 (06) \cr
\deepstrut$~4\cdot ~6^3$ && 5.740 && 0.47735 (0653) && 2.74\ $10^{-2}$ 
&& 3.06 (014)\ $10^{-2}$ && 16 \% && 5.6940 (05) \cr
\noalign{\vskip-3.5pt}
$~4\cdot ~8^3$ && 5.600 && 0.09220 (0114) && 2.59\ $10^{-3}$
&& none         && &&        \cr
$~4\cdot ~8^3$ && 5.670 && 0.22317 (0830) && 1.80\ $10^{-2}$ 
&& 2.33 (010)\ $10^{-2}$ && 11 \% && 5.6915 (35) \cr
$~4\cdot ~8^3$ && 5.693 && 0.32431 (1003) && 2.50\ $10^{-2}$ 
&& 2.50 (008)\ $10^{-2}$ && 12 \% && 5.6937 (22) \cr
\deepstrut$~4\cdot ~8^3$ && 5.720 && 0.45422 (0555) && 1.61\ $10^{-2}$ 
&& 2.12 (012)\ $10^{-2}$ && 23 \% && 5.6922 (51) \cr
\noalign{\vskip-3.5pt}
$~4\cdot 10^3$ && 5.600 && 0.06563 (0066) && 1.20\ $10^{-3}$ 
&& none         && && \cr
$~4\cdot 10^3$ && 5.680 && 0.21619 (0834) && 1.72\ $10^{-2}$
&& 2.047 (093)\ $10^{-2}$&& 12 \% && 5.6901 (17) \cr
$~4\cdot 10^3$ && 5.693 && 0.32961 (0900) && 1.86\ $10^{-2}$ 
&& 1.984 (069)\ $10^{-2}$&& 17 \% && 5.6867 (18) \cr
\deepstrut$~4\cdot 10^3$ && 5.710 && 0.41699 (0815) && 1.33\ $10^{-2}$ 
&& 1.997 (160)\ $10^{-2}$&& 30 \% && 5.6922 (28) \cr
\noalign{\vskip-3.5pt}
$~4\cdot 12^3$ && 5.620 && 0.05761 (0052) && 0.10\ $10^{-2}$ 
&& none         && && \cr
$~4\cdot 12^3$ && 5.681 && 0.15435 (1002) && 1.16\ $10^{-2}$ 
&& 2.04 (025)\ $10^{-2}$ && 28~\% && 5.6925 (17) \cr
$~4\cdot 12^3$ && 5.691 && 0.27161 (1581) && 1.92\ $10^{-2}$ 
&& 1.95 (009)\ $10^{-2}$ && 30~\% && 5.6903 (16) \cr
\deepstrut$~4\cdot 12^3$ && 5.703 && 0.40754 (0718) && 0.87\ $10^{-2}$ 
&& none         && &&        \cr
\noalign{\vskip-3.5pt}
$~4\cdot 14^3$ && 5.682 && 0.15595 (1196) && 1.14\ $10^{-2}$ 
&& 1.64 (011)\ $10^{-2}$ && 30 \% && 5.6884 (15) \cr
$~4\cdot 14^3$ && 5.691 && 0.24181 (1895) && 1.82\ $10^{-2}$ 
&& 1.87 (012)\ $10^{-2}$ && 31 \% && 5.6920 (12) \cr
\deepstrut$~4\cdot 14^3$ && 5.698 && 0.37534 (0768) && 0.73\ $10^{-2}$ 
&& none         && &&        \cr
\noalign{\vskip-3.5pt}
$~4\cdot 16^3$ && 5.683 && 0.10277 (1031) && 0.68\ $10^{-2}$ 
&& 1.78 (009)\ $10^{-2}$ &&102 \% && 5.6902 (15) \cr
$~4\cdot 16^3$ && 5.691 && 0.22025 (1992) && 1.72\ $10^{-2}$ 
&& 1.76 (012)\ $10^{-2}$ && 33 \% && 5.6915 (09) \cr
$~4\cdot 16^3$ && 5.692 && 0.25229 (1937) && 1.73\ $10^{-2}$ 
&& 1.78 (013)\ $10^{-2}$ && 30 \% && 5.6914 (09) \cr
\deepstrut$~4\cdot 16^3$ && 5.697 && 0.36245 (0793) && 0.68\ $10^{-2}$ 
&& none         && &&     \cr
\noalign{\vskip-3.5pt}
$~4\cdot 20^3$ && 5.690 && 0.16688 (1729) && 1.41\ $10^{-2}$ 
&& 1.65 (009)\ $10^{-2}$ && 44 \% && 5.6914 (05) \cr
$~4\cdot 20^3$ && 5.691 && 0.20443 (2119) && 1.71\ $10^{-2}$ 
&& 1.76 (008)\ $10^{-2}$ && 55 \% && 5.6913 (05) \cr
\deepstrut$~4\cdot 20^3$ && 5.692 && 0.19617 (2420) && 1.62\ $10^{-2}$ 
&& 1.73 (012)\ $10^{-2}$ && 52 \% && 5.6926 (06) \cr
\noalign{\vskip-3.5pt}
$~4\cdot 24^3$ && 5.691 && 0.08825 (1466) && 0.73\ $10^{-2}$
&& 2.43 (085)\ $10^{-2}$ && 50 \% && 5.6933 (09)\cr
$~4\cdot 24^3$ && 5.693 && 0.31780 (1379) && 0.76\ $10^{-2}$ 
&& 1.55 (033)\ $10^{-2}$ && 32 \% && 5.6909 (08)\cr
}}}   
\smallskip
\vtop{\noindent\tenpoint
All error bars are calculated with respect to twenty jackknife bins
and corrected for the bias. For the maximum of the 
susceptibility, the
bias correction is given in percent of the error bar. The result ``none'' 
means that maximum of the susceptibility is either unreliable or out of 
the controled $\beta$-range.}
\bigskip
\endinsert 

\midinsert 
\centerline{\bf Table 11: Polyakov Susceptibility by 
                         standard error propoagation.}
$$
\vbox{\tabskip=0pt\offinterlineskip\tenpoint
   \halign{\strut\hfil#\hfil\tabskip=.7em&
\vrule#&
     \hfil#\hfil&
\vrule#&
\hfil#\hfil&
\vrule#&
\hfil#\hfil
          \tabskip=0pt\cr
\tablerule
\strutrule
$L_t\cdot L^3$&&
\omit{\hfil $L^{-3}\chi_{\max}(P) $, ~goodness \hfil}&&
\omit{\hfil $\beta_{\max}$, ~goodness       \hfil}&&
\omit{\hfil \# \hfil}\cr
\tablerule
\omit\smallstrut&&&&&&\cr
\noalign{\vskip-3pt}
$~2\cdot 12^3$ && 1.95 (05)\ $10^{-1}$, ~0.83 && 5.0926 (06), ~0.67 && 2 \cr
               &&                             &&                    &&   \cr
$~4\cdot ~6^3$ && 2.92 (05)\ $10^{-2}$, ~0.49 && 5.6957 (26), ~0.97 && 4 \cr
$~4\cdot ~8^3$ && 2.37 (06)\ $10^{-2}$, ~0.03 && 5.6930 (18), ~0.86 && 3 \cr
$~4\cdot 10^3$ && 2.01 (06)\ $10^{-2}$, ~0.86 && 5.6891 (12), ~0.19 && 3 \cr
$~4\cdot 12^3$ && 1.96 (09)\ $10^{-2}$, ~0.74 && 5.6913 (12), ~0.35 && 2 \cr
$~4\cdot 14^3$ && 1.75 (09)\ $10^{-2}$, ~0.17 && 5.6906 (10), ~0.07 && 2 \cr
$~4\cdot 16^3$ && 1.77 (07)\ $10^{-2}$, ~0.99 && 5.6913 (06), ~0.74 && 3 \cr
$~4\cdot 20^3$ && 1.72 (06)\ $10^{-2}$, ~0.65 && 5.6917 (04), ~0.20 && 3 \cr
$~4\cdot 24^3$ && 1.67 (31)\ $10^{-2}$, ~0.34 && 5.6920 (06), ~0.05 && 2 \cr
}   
}   
$$
\vtop{\tenpoint\noindent
Results are calculated from table 10 according to standard error progagation.
The last column shows how many data sets were combined.}
\endinsert 

\midinsert 
\centerline{\bf Table 12:  Patching for the Polyakov Susceptibility.}
$$
\vbox{\tabskip=0pt\offinterlineskip\tenpoint
   \halign{\strut\hfil#\hfil\tabskip=.7em&\vrule#&
     \hfil#\hfil&\vrule#&\hfil#\hfil&\vrule#&#\hfil
          \tabskip=0pt\cr
\tablerule
\strutrule
$L_t\cdot L^3$&&
\omit{\hfil $L^{-3}_{\chi \max}(P)$\hfil}&&
\omit{\hfil $\beta_{\max}$\hfil}&&
\omit{\hfil \# Patches: weights\hfil}\cr
\tablerule
\omit\smallstrut&&&&&&\cr
\noalign{\vskip-3pt}
$~2\cdot 12^3$ && 1.95 (06)\ $10^{-1}$ && 5.0927 (06) && 2: 0.68, 0.32.\cr
&&&&&&\cr
$~4\cdot ~6^3$ && 2.95 (04)\ $10^{-2}$ && 5.7007 (26) && 
4: 0.14, 0.36, 0.33, 0.17\cr
$~4\cdot ~8^3$ && 2.37 (06)\ $10^{-2}$ && 5.6922 (12) && 
3: 0.37, 0.42, 0.21.\cr
$~4\cdot 10^3$ && 2.00 (05)\ $10^{-2}$ && 5.6885 (12) && 
3: 0.41, 0.47, 0.12.\cr
$~4\cdot 12^3$ && 1.95 (08)\ $10^{-2}$ && 5.6910 (13) && 2: 0.31, 0.69.\cr
$~4\cdot 14^3$ && 1.71 (06)\ $10^{-2}$ && 5.6896 (07) && 
3: 0.40, 0.41, 0.19.\cr
$~4\cdot 16^3$ && 1.73 (06)\ $10^{-2}$ && 5.6910 (05) && 
4: 0.15, 0.36, 0.36, 0.13.\cr
$~4\cdot 20^3$ && 1.70 (05)\ $10^{-2}$ && 5.6917 (04) && 
3: 0.39, 0.34, 0.27.\cr
$~4\cdot 24^3$ && 1.88 (09)\ $10^{-2}$ && 5.6919 (12) && 2: 0.46, 0.544.\cr
}   
}   
$$
\vtop{\tenpoint\noindent
The last column gives the number of patched data sets and the relative
weights, ordered by increasing $\beta_0$. All error bars are 
calculated with respect to twenty jackknife bins
and corrected for the bias.} 
\endinsert 


\midinsert 
\centerline{\bf Table 13: $L_{t}=4$ FSS fits to $\chi_{\max}(P)$}
$$
\vbox{\tabskip=0pt\offinterlineskip\tenpoint
   \halign{\strut\hfil#\hfil\tabskip=.7em&
\vrule#&
\hfil#\hfil&
\hfil#\hfil&
\hfil#\hfil&
\vrule#&
\hfil#\hfil&
\hfil#\hfil&
\hfil#\hfil
\tabskip=0pt\cr
\tablerule
\strutrule
&&
\multispan3{\hfil $(Standard\;\;error\;\;propagation)$\hfil}&&
\multispan3{\hfil $(Patching)$\hfil}\cr
\noalign{\vskip-2pt}
$L$ range&&
\omit{\hfil $10^2a_{1}$\hfil}&
\omit{\hfil $a_{2}$\hfil}&
\omit{\hfil $Q$\hfil}&&
\omit{\hfil $10^2a_{1}$\hfil}&
\omit{\hfil $a_{2}$\hfil}&
\omit{\hfil $Q$\hfil}\cr
\tablerule
\omit\smallstrut&&&&&&&&\cr
\noalign{\vskip-3pt}
~6 - 14&& 1.78(05)     &  2.5(2.0)      & 0.17     &&
                1.73(4)     &~2.7(2.0)      &0.04\cr
~6 - 16 &&  1.76(04)      &   2.6(2.0)       &  0.21      &&
                 1.71(4)      &~2.7(2.0)       & 0.05  \cr
~6 - 24 &&  1.74(04)      &   2.6(2.0)       &  0.33      &&
                 1.71(3)      &~2.7(2.0)       & 0.04  \cr
~8 - 16 &&  1.68(06)      &   3.5(5.0)       &  0.73      &&
                 1.63(5)      &~3.8(5.0)       & 0.46  \cr
~8 - 20 &&  1.68(04)      &   3.5(4.0)       &  0.87      &&
                 1.63(4)      &~3.8(4.0)       & 0.63  \cr
~8 - 24 &&  1.68(05)      &   3.5(5.0)       &  0.94      &&
                 1.66(4)      &~3.5(4.0)       & 0.16  \cr
 10 - 24 &&  1.68(05)      &   3.4(9.0)       &  0.87      &&
                 1.68(5)      &~3.2(8.0)       & 0.11   \cr
 12 - 24 &&  1.65(08)      &   4.8(2.3)     &  0.84      &&
                 1.69(7)      &~2.8(1.9)     & 0.06   \cr
 14 - 24 &&  1.70(10)     &   1.8(4.3)     &  0.91      &&
                 1.79(8)      &-2.5(3.0)     & 0.25   \cr}}$$  
\vtop{\tenpoint\noindent
Eq. (6.2) is used to fit the $\chi_{max}(P)$ in tables 11 resp. 12.}
\endinsert 

\medskip
\heading{POLYAKOV LINE SUSCEPTIBILITY}

Refs. [15,31] have advocated the FSS analysis of the susceptibility of 
the projected Polyakov line as a good indicator of the order of the
$SU(3)$ phase transition. Since we have measured and recorded
the real and imaginary parts of the Polyakov line along with the
plaquette action, for each of our runs, we can apply the
procedures discussed in Section 4 to the 
spectral-density FSS analysis of this quantity. 
\midinsert 
\centerline{\bf Table 14: $L_{t}=4~\chi(P)$ 
FSS fits of $\beta_{\max}(L)$}
$$
\vbox{\tabskip=0pt\offinterlineskip\tenpoint
   \halign{\strut\hfil#\hfil\tabskip=.7em&
\vrule#&
\hfil#\hfil&
\hfil#\hfil&
\hfil#\hfil&
\vrule#&
\hfil#\hfil&
\hfil#\hfil&
\hfil#\hfil
\tabskip=0pt\cr
\tablerule
\strutrule
&&
\multispan3{\hfil $(Standard\;\;error\;\;propagation)$\hfil}&&
\multispan3{\hfil $(Patching)$\hfil}\cr
\noalign{\vskip-2pt}
$L$ range&&
\omit{\hfil $\beta_c$\hfil}&
\omit{\hfil $a$\hfil}&
\omit{\hfil $Q$\hfil}&&
\omit{\hfil $\beta_c$\hfil}&
\omit{\hfil $a$\hfil}&
\omit{\hfil $Q$\hfil}\cr
\tablerule
\omit\smallstrut&&&&&&&&\cr
\noalign{\vskip-3pt}
~6 - 24&& 5.6914(3) & ~0.4(5.0)    &0.20     &&
                5.6907(3)  &~1.1(5.0)    & 0.005     \cr
~8 - 24&& 5.6917(4)  &-0.7(8.0)     &0.31     &&
                5.6910(4)  & -0.2(7.0)     & 0.060    \cr
10 - 24&& 5.6921(4)  & -2.8(1.3)&   0.94     &&
                5.6918(4)  & -3.4(1.3)   & 0.600     \cr
12 - 24&& 5.6920(5)  & -2.3(2.2)   &0.87     &&
                5.6918(6)  & -3.9(2.3)   & 0.440     \cr
14 - 24&& 5.6923(6)  & -4.4(3.3)   &0.98     &&
                5.6923(7)  & -6.7(3.0)   & 0.740     \cr}}$$
\vtop{\tenpoint\noindent
Eq. (4.7) is used to fit the $\beta_{max}$ in tables 11 resp. 12.}
\endinsert 

\midinsert 
\centerline{\bf Table 15: $L_{t}=2$ FSS fits of $\chi_{\max}(P)$ 
 and its $\beta_{\max}(L)$}
$$
\vbox{\tabskip=0pt\offinterlineskip\tenpoint
   \halign{\strut\hfil#\hfil\tabskip=.7em&
\vrule#&
\hfil#\hfil&
\hfil#\hfil&
\hfil#\hfil&
\vrule#&
\hfil#\hfil&
\hfil#\hfil&
\hfil#\hfil
\tabskip=0pt\cr
\tablerule
\strutrule
$L$ range&&
\omit{\hfil $10^2a_{1}$\hfil}&
\omit{\hfil $a_{2}$\hfil}&
\omit{\hfil $Q$\hfil}&&
\omit{\hfil $\beta_c$\hfil}&
\omit{\hfil $a$\hfil}&
\omit{\hfil $Q$\hfil}\cr
\tablerule
\omit\smallstrut&&&&&&&&\cr
\noalign{\vskip-3pt}
~6 - 12&&18.8(5.0) & -08.3(01.3) &0.0007&&5.0942(08)&-2.29 (56.0)&0.53\cr
~8 - 12&&21.2(8.0)&-26.0(05.0) &0.7100&&5.0942(10)&-2.31(97.0)& 0.26\cr
10 - 12&&20.6(1.7)&-19.0(20.0)& - &&5.0915(26)&~2.10(04.1)& -\cr}}$$
\vtop{\tenpoint\noindent
The susceptibility maxima for $L_t=2$ are fitted to eq. (6.2) and the
corresponding $\beta_{max}$ to eq. (4.7).}
\endinsert 

We have a time series of measurements of 
the lattice average of the Polyakov line in the Euclidean time direction,
$\Omega = V^{-1} \sum_x \Omega_x$. These are complex numbers:
$\Omega={\rm Re}~\Omega+~{\bf i}{\rm Im}~\Omega= \rho {\rm e}^{i\phi}$.
(In our definition, $\Omega_x$ is larger than in Ref. [15] by the colour
factor 3).

The projected real part $P$ is computed as

$$ \cases{1.\ ~P = {\rm Re}~\Omega\ ~{\rm for\ \phi \in [-\pi/3,\pi/3),} \cr
       2.\ ~P  = {\rm Re}~
          \exp(-i2\pi/3) \Omega\ ~{\rm for\ \phi \in [\pi/3,\pi),} \cr
       3.\ ~P = {\rm Re}~
          \exp(i2\pi/3) \Omega\ ~{\rm for\ \phi \in [-\pi,-\pi/3).}}
    \eqno(6.1) 
$$ 

$P$ provides a clear distinction between the $Z_3$-symmetric phase of
the theory (where it is close to zero) and any of the three broken phases
(where it is nonzero). Since $\phi$ is projected out, there is no
cancellation due to tunneling between the three broken phases
(which makes the time series average 
$\bar\Omega$ vanish on a finite lattice, even above deconfinement).
The susceptibility (variance) of $P$ can now be analyzed
exactly like the variance of the action was analyzed in Section 4.
(To compute the moments of $P$, we calculate the Boltzmann factors 
from the corresponding action measurements.)
The results corresponding to table 2 are collected in table 10.
(Remember that our susceptibilities differ from those of Ref. [15]
by a factor 9, because of the normalization of $\Omega_x$.)

Table 11 shows the results of the standard error propagation 
analysis, as applied to the valid results in table 10. Since
the $4\cdot 4^3$ lattices yield no valid results, there is no  $4\cdot 4^3$
entry in table 11.
Like for the specific heat, the data sets for $L=14,24$ ($L_t=4$)
give poor estimates of $\beta_{max}$ and the error
on $L^{-3}\chi_{max}$ is large for $L=24$.
Again, patching with weights given by eq. (2.9b)
(where we put $f_i=P_i$) improves the situation, as can be seen
in table 12. However, we notice that the patched 
$L^{-3}\chi_{max}$ for $L=24$ ($L_t=4$) violates the trend in the
results for the other $L$.

Regarding the FSS behavior of the susceptibility maxima and of the
corresponding values of $\beta$, the same considerations apply as 
discussed in Section 4. Assuming a first order transition, we fit
the data in tables 11 and 12 with the analog of eq. (4.4b):

$$ L^{-3}\chi(P)\ =\ a_1 + a_2 L^{-3}
   \eqno(6.2) 
$$

Tables 13 and 15 show the results of these fits for $L_t=4$
and $L_t=2$ respectively. For $L_t=4$, 
we see that the patching data (which are of better
quality) are more restrictive than those obtained by error propagation.
Our suspicion about the data $L=24$ (presumably insufficient statistics) 
are confirmed by the $Q$ values for the corresponding fits. In addition, 
$L=6$ is presumably a too small lattice. The best estimate of the order
parameter jump $ \triangle P\  =\ 2 \sqrt{a_1}$ should be the one
obtained from the patched fit $L=8-24$ ($Q$ is still acceptable, and
while it makes sense to ignore the $L=6$ result we cannot discard our
largest lattice $L=24$ even though the statistics is poor.)
We get

$$ 
\triangle P\  =\  (0.26\pm 0.04) , ~~~(L_t=4).   \eqno(6.3) 
$$
For $L_t=2$ too, the $L=6$ data spoil the first order fit;
the best estimate is

$$ 
\triangle P\  =\  (0.92\pm 0.18) , ~~~(L_t=2)   \eqno(6.4)
$$
from the range $L=8 - 12$.

We can also use the $\beta$ values corresponding to the maxima of 
the susceptibility, in order to obtain $\beta_c$ through the fit (4.7). 
The results of this exercise appear in tables 14 (for $L_t=4$) and 15
(for $L_t=2$). Our best estimates for $\beta_c^{~\rm susc}$ are

$$ 
\beta_c^{~\rm susc}\ =\ 5.6918 \pm 0.0004 ~~~(L_t=4)  \eqno(6.5) 
$$
(using $L=10-24$ by patching), and

$$ 
\beta_c^{~\rm susc}\ =\ 5.0942 \pm 0.0008   ~~~(L_t=2)  \eqno(6.6) 
$$
using $L=6-12$. Note that the $\beta_c$ fit selects optimal ranges which 
are different from the one we used for the fit to the susceptibility maxima. 
Obviously, this is allowed in principle since the height of the peak and 
its location in $\beta$ may be independent functions of $L$.

While $\beta_c^{~\rm susc}$ for $L_t=2$ is seen to be consistent with those
estimated from the analysis of the specific heat (cf. eq. (4.11)) and of 
the partition function zeros (cf. eq. (5.15)), $\beta_c^{~\rm susc}$ for 
$L_t=4$ is rather small and becomes only consistent on a two $\sigma$ level
(cf. eqs. (4.10), (5.13)). 
This may indicate that with our statistics (presumably due to long time
correlations) the Polyakov susceptibility is not accurate enough to 
allow really precise fits.
 
\medskip
\heading{SUMMARY AND CONCLUSIONS}

Spectral density methods greatly facilitate accurate FSS calculations.
One can calculate pseudocritical couplings precisely
and extrapolate towards the infinite volume critical coupling.
From the specific heat, the analysis of the
partition function zeros and the Polyakov loop susceptibilities we
have obtained three estimates of the infinite volume critical $\beta$
which differ somewhat due to remaining systematic errors. In the 
absence of other strong criteria, one may average these estimates, weighted
by their error bars and quote as the final error the best error bar of the
single estimates (one can not use error propagation as all results are
obtained from the same configurations). In this way we obtain from
(4.9), (5.13) and (6.3)

$$ 
\beta_c\ =\ 5.6927 \pm 0.0004 , ~~~(L_t=4),  \eqno(7.1) 
$$
and from (4.1), (5.15) and (6.4):

$$ 
\beta_c\ =\ 5.0933 \pm 0.0007 , ~~~(L_t=2),  \eqno(7.1) 
$$
The achieved accuracy improves estimates from the pioneering literature
~[12] by one order of magnitude in the error bar 
(translating to two orders of magnitude in computer time).

Another notable result are the latent heats (4.8) and (4.11), which
are consistent with independent results by other groups~[15,32].
Again the spectral density FSS approach works quite well. The possibility
to calculate a latent heat and, similarly, an order parameter jump
self-consistently is in our opinion the strongest argument
in favour of the first order nature of this phase transition. Whereas
for $L_t=2$ one observes, additionally, a clear double peak structure
the $L_t=4$ transition is fairly weak and a double peak structure begins
(marginally) only to develop from $L=16$ on. It is remarkable that the
FSS behavior is nevertheless already quite indicative for the first
order nature. This seems to be driven by the increase of the width of 
the almost Gaussian looking action density distribution.

For $L_t=4$ the analysis of the partition function zeros has turned
out to be more subtle than previously anticipated. Nevertheless
from $L=14$ on the results seem to be conclusive and the obtained
estimate (5.12) of the critical exponent $\nu$  is consistent with
$\nu = 1/d$ ($d=3$) and rounds the picture of a weak first order
transition. For $L_t=2$ the same analysis is fairly unproblematic
and the results of [21] are nicely improved and confirmed. The $L_t=2$
estimate (5.14) for $\nu$ is, of course, consistent with a first order
transition.
\bigskip

\noindent{\bf ACKNOWLEDGEMENTS} 
\medskip
The MC data were produced on Florida State University's ETA$^{10}$'s. In
addition this work made heavy use of the FSU Physics Department HEP VAX and
the Supercomputer Computations Research Institute RISC clusters and their
attached storage facilities. This research project was partially funded by
the U.S. National Science Foundation under grant INT-8922411 and by the the
U.S. Department of Energy under contracts DE-FG05-87ER40319 and
DE-FC05-85ER2500. Nelson Alves is supported by the German Humboldt
foundation. 

\bigskip
\noindent{\bf REFERENCES}   
\medskip
\parskip=4pt plus2pt minus1pt
\item{1)} Z.W. Salsburg, J.D. Jackson, W. Fickett and W.W. Wood,
          {\it J. Chem. Phys.}, {\bf 30}, (1959), 65.

\item{2)} I.R. McDonald and K. Singer, Discuss. {\it Faraday Soc.}, 
         {\bf 43}, (1967), 40.

\item{3)} J.P. Valleau and D.N. Card, {\it J. Chem. Phys.}, {\bf 57}, 
         (1972), 5457.

\item{4)} C.H. Bennett, {\it J. Comp. Phys.}, {\bf 22}, (1976), 245.

\item{5)} M. Falcioni, E. Marinari, M.L. Paciello, G. Parisi and
          B. Taglienti, {\it Phys. Lett.}, {\bf 108B}, (1982), 331;
          E.~Marinari, {\it Nucl. Phys.}, {\bf B235}, [FS11], (1984), 123.

\item{6)} G. Bhanot, S. Black, P. Carter and R. Salvador, 
          {\it Phys. Lett.}, {\bf B183}, (1987), 331.

\item{7)} A.M. Ferrenberg and R.H. Swendsen, {\it Phys. Rev. Lett.},
          {\bf 61}, (1988), 2635, [erratum 63, (1989), 1658].

\item{8)} A.M. Ferrenberg and R.H. Swendsen, {\it Phys. Rev. Lett.}, 
           63, (1989), 1196.

\item{9)} N.A. Alves, B.A. Berg and R. Villanova, {\it Phys. Rev.}, 
{\bf B41}, (1990), 383.
 
\item{10)} N.A. Alves, B.A. Berg and R. Villanova, {\it Phys. Rev.}, 
{\bf B43}, (1991), 5846.

\item{11)} S. Aoki {\it et al.} (The Teraflop Collaboration):
         {\it The QCD Teraflop Project}, to be published
          in {\it Int. J. Mod. Phys. C.}

\item{12)} J. Kogut {\it et al}., {\it Phys. Rev. Lett.}, {\bf 51}, 
          (1983); T. Celik, J.~Engels
          and H.~Satz, {\it Phys. Lett.}, {\bf 129B}, (1983), 323; 
          A.~Kennedy {\it et al}.,
          {\it Phys. Rev. Lett.}, {\bf 54}, (1985), 87; N. Christ and 
       A. Terrano, {\it Phys. Rev. Lett.}, {\bf 56}, (1986), 111.

\item{13)} P. Bacilieri {\it et. al.}, {\it Phys. Rev. Lett.}, {\bf 61}, 
         (1988), 1545; {\it Phys. Lett.}, {\bf 224B}, (1989), 333.

\item{14)} A. Ukawa, Proceedings of the LAT89 Capri conference, 
           {\it Nucl. Phys. B}, (Proc. Suppl.) {\bf 17}, (1990), 118.

\item{15)} M. Fukugita, M. Okawa and A. Ukawa, {\it Nucl. Phys.}, 
          {\bf B337}, (1990), 181.

\item{16)} B.A. Berg, R. Villanova and C. Vohwinkel, {\it Phys. Rev. 
Lett.}, 62, (1989), 2433.

\item{17)} N.A. Alves, B.A. Berg and S. Sanielevici, {\it Phys. Rev. 
Lett.}, {\bf 64}, (1990), 3107.

\item{18)} A.D. Sokal, `` Monte Carlo Methods in Statistical 
           Mechanics: Foundations and New Algorithms'', preprint, 
           New York University (1989).

\item{19)} B.A. Berg and J. Stehr, Z. {\it Phys}, {\bf C9}, (1981), 333.

\item{20)} H. Flyvbjerg and H.G. Peterson, {\it J. Chem. Phys.}, {\bf 91}, 
           (1989), 461.

\item{21)} M. Karliner, S. Sharpe and Y. Chang, {\it Nucl. Phys.}, 
{\bf B302}, (1988), 204.

\item{22)} B.A. Berg, A. Devoto and C. Vohwinkel, {\it Comp. Phys. 
Commun.}, {\bf 51}, (1988), 331.

\item{23)} B.A. Berg, preprint, SCRI 90-100, to be published in 
           {\it Comp. Phys. Commun.}

\item{24)} M.E. Fisher, in Nobel Symposium 24, B. Lundquist and 
           S.~Lundquist (editors), Academic Press, New York, 1974.

\item{25)} B.L. Van der Waerden, {\it Mathematical Statistics} (Springer,
           New York, 1969).

\item{26)} W. Press et al., {\it Numerical Recipes}, Cambridge University
           Press, London, 1986.

\item{27)} C. Itzykson, R.B. Pearson and J.B. Zuber, {\it Nucl. Phys.},
           {\bf B220}, [FS8], (1983), 415.

\item{28)} G. Bhanot, S. Black, P. Carter and R. Salvador, {\it Phys. 
Lett.}, {\bf B183}, (1987), 331.

\item{29)} K. Bitar, {\it Nucl. Phys.}, B300, [FS22], (1988), 61.

\item{30)} N.A. Alves, B.A. Berg and S. Sanielevici, 
           {\it Phys. Lett.}, {\bf 241B}, (1990), 557.

\item{31)} M. Fukugita, H. Mino, M. Okawa and A. Ukawa, {\it Nucl. Phys. 
B}, (Proc. Suppl.), {\bf 20}, (1991), 258.

\item{32)} F. Brown, N. Christ, Y. Deng, M. Gao and T. Woch,
           {\it Phys. Rev. Lett.}, {\bf 61}, (1988), 2058.

\bye


%
%
%
%
%
\input fontdefs
\input 14pt.font
\twelvepoint
\overfullrule=0pt   
\baselineskip=12.4pt
\hsize=6truein
\vsize=8.5truein
\hoffset=6.5in			
   \advance\hoffset by-\hsize	
   \divide\hoffset by2		
\voffset=9in			
   \advance\voffset by-\vsize
   \divide\voffset by2
\def\resetpage{\global\pageno=1}
\clubpenalty=10000
\widowpenalty=10000
\parskip=1.5ex plus.5ex minus.5ex
\parindent=2em
\def\boxit#1{\vbox{\hrule height 0pt\hbox{\vrule width0pt%
	\vbox{\hbox{#1}}\vrule width0pt}\hrule height 0pt}}
\def\today{\ifcase\month\or
  January\or February\or March\or April\or May\or June\or
  July\or August\or September\or October\or November\or December\fi
  \space\number\day, \number\year}

\footline={\ifnum\pageno=1\hfill\else\hfil\folio\hfil\fi}

\newdimen\cboxindent  
\newdimen\cboxflex    
\cboxindent=3.5cm
\cboxflex=5cm
\def\cbox#1{\vbox
   {\advance\leftskip by\cboxindent plus\cboxflex minus\cboxflex
    \advance\rightskip by\cboxindent plus\cboxflex minus\cboxflex
    \parindent=0pt
    \parfillskip=0pt
\hyphenpenalty=10000
    #1}}
\def\titlecbox#1{\vbox to 94.85truept
   {\advance\leftskip by\cboxindent plus\cboxflex minus\cboxflex
    \advance\rightskip by\cboxindent plus\cboxflex minus\cboxflex
    \parindent=0pt
    \parfillskip=0pt
\hyphenpenalty=10000
    \vfill#1\vfill}}
\def\authorcbox#1{\vbox to 72.27truept
   {\advance\leftskip by\cboxindent plus\cboxflex minus\cboxflex
    \advance\rightskip by\cboxindent plus\cboxflex minus\cboxflex
    \parindent=0pt
    \parfillskip=0pt
\hyphenpenalty=10000
    \vfill#1\vfill}}
\def\extracbox#1{\vbox to 63.24truept
   {\advance\leftskip by\cboxindent plus\cboxflex minus\cboxflex
    \advance\rightskip by\cboxindent plus\cboxflex minus\cboxflex
    \parindent=0pt
    \parfillskip=0pt
\hyphenpenalty=10000
    #1\vfill}}
%
%
%

\newbox\covernum
\newbox\covertitlebox
\newbox\coverauthorbox
\newbox\coverextrabox
\newbox\coveratype
\newbox\coverptype
\newbox\coverdatebox

\def\covertitle#1{\global\cboxindent=1.5cm\global\cboxflex=1cm\relax%
	\setbox\covertitlebox=\titlecbox{\fourteenpoint\bf #1}}
\def\coverauthor#1{\global\cboxindent=3.5cm\global\cboxflex=3cm\relax%
	\setbox\coverauthorbox=\authorcbox{\fourteenpoint\bf #1}}
\def\coverpreprint#1{\global\setbox\covernum=\hbox{\fourteenpoint\bf #1}}
\def\coverextra#1{\global\cboxindent=3.5cm\global\cboxflex=3cm\relax%
	\setbox\coverextrabox=\extracbox{\fourteenpoint\bf{(#1)}}}
\def\coverdate#1{\global\setbox\coverdatebox=\hbox{\fourteenpoint\bf #1}}
\def\authortype#1{\global\setbox\coveratype=\hbox{\fourteenpoint\bf #1}}
\def\papertype#1{\global\setbox\coverptype=\hbox{\fourteenpoint\bf #1}}
\def\makecover{%
\ifvoid\coverextrabox%
	\global\setbox\coverextrabox=\vtop to 63.24truept{\hfill}%
\fi%
\hsize=6.5truein
\vsize=9truein
\hoffset=6.5in			
   \advance\hoffset by-\hsize	
   \divide\hoffset by2		
\voffset=9in			
   \advance\voffset by-\vsize
  \divide\voffset by2
\resetpage
\input disclaimer
\resetpage
\boxit{\vtop to \vsize{\fourteenpoint\bf\null
\vskip2.2in
\centerline{\boxit{\vtop{\copy\covertitlebox}}}
\vskip 1truecm
\boxit{\centerline{by}}
\vskip 1truecm
\centerline{\boxit{\vtop{\copy\coverauthorbox}}}
\vskip 1truecm
\centerline{\boxit{%
	\cbox{FSU-SCRI-91\box\coveratype\box\coverptype-\box\covernum}}}
\smallskip
\centerline{\boxit{\vtop{\box\coverextrabox}}}
\vskip9truept
\ifvoid\coverdatebox
	\centerline{\boxit{\today}}
\else
	\centerline{\boxit{\box\coverdatebox}}
\fi
\vfill}}
\eject}

\covertitle{SPECTRAL DENSITY STUDY OF THE SU(3) DECONFINING PHASE TRANSITION}
\coverauthor{Nelson~A.~Alves, Bernd~A.~Berg and Sergiu~Sanielevici}
\coverpreprint{93}
\coverdate{July 1991}
                             
\makecover
\bye